\documentclass[journal]{IEEEtran}
\usepackage{cite}

\ifCLASSINFOpdf
  \usepackage[pdftex]{graphicx}
\else
   \usepackage[dvips]{graphicx}
\fi
\usepackage{array}

\usepackage[tight,footnotesize]{subfigure}
\begin{document}
%
\title{The Recursive Form of Error Bounds for RFS State and Observation with $P_d<1$}
%
%
%

\author{Huisi~Tong,~\IEEEmembership{Student Member,~IEEE,}
        Hao~Zhang,~\IEEEmembership{Member,~IEEE,}
        Huadong~Meng,~\IEEEmembership{Member,~IEEE,}
        and~Xiqing~Wang,~\IEEEmembership{Member,~IEEE}
\thanks{The authors are with the Department
of Electrical Engineering, Tsinghua University, Beijing, China, e-mail: tonghs08@mails.tsinghua.edu.cn}
\thanks{This work is supported by $?$}
\thanks{Manuscript received $?$; revised $?$.}}

%
%

\markboth{Journal of \LaTeX\ Class Files,~Vol.~?, No.~?, ?~?}%
{Shell \MakeLowercase{\textit{et al.}}: Bare Demo of IEEEtran.cls for Journals}
%



\maketitle

\begin{abstract}
In the target tracking and its engineering applications, recursive
state estimation of the target is of fundamental importance. This
paper presents a recursive performance bound for dynamic estimation
and filtering problem, in the framework of the finite set statistics
for the first time. The number of tracking algorithms with
set-valued observations and state of targets is increased sharply
recently. Nevertheless, the bound for these algorithms has not been
fully discussed. Treating the measurement as set, this bound can be
applied when the probability of detection is less than unity.
Moreover, the state is treated as set, which is singleton or empty
with certain probability and accounts for the appearance and the
disappearance of the targets. When the existence of the target state
is certain, our bound is as same as the most accurate results of the
bound with probability of detection is less than unity in the
framework of random vector statistics. When the uncertainty is taken
into account, both linear and non-linear applications are presented
to confirm the theory and reveal this bound is more general than
previous bounds in the framework of random vector statistics.In
fact, the collection of such measurements could be treated as a
random finite set (RFS).
\end{abstract}


%
\IEEEpeerreviewmaketitle

\section{Introduction}\label{s1}
%
%
%
%
\IEEEPARstart{I}{n} the target tracking and its engineering
applications, recursive state estimation of the target is of
fundamental importance \cite{Ref1}. However, the tracking system may
not receive the information of the target, which is result from the
probability of detection $P_d<1$. Moreover, even though a
measurement is received by the tracking system, it is hard to
determine whether it is produced by the target or not, when the
false alarm $P_{FA}>0$. Therefore, at any time step, the number of
measurement is random and it is unlikely to know whether there is
missing detection or there are false measurements \cite{Ref2},
\cite{Ref3}.

This paper presents a recursive performance bound of estimation
error with set-valued observations and state of targets, which is
more general than the one with random vector measurement and state.
The estimation problem where the both the measurement and the state
are finite set, is very important in defense and surveillance
\cite{Ref2}, \cite{Ref3}. The reason is that we cannot determine
whether the target exists or not from the measurement and meanwhile
the existence of the target varies with the time passing. In fact,
the bound in the framework of random vector statistics is only a
special case of our bound, when the existence of the target state is
certain. Therefore, this bound is a limit of a dynamic estimation
error for the problem that the state set of the target is a Markov
process, the measurement set is statistically dependent on the
existence of the target, and the number of the points in the state
set as well as measurement set is random at any time step.

The error in this paper is a distance between the state set and the estimation set and thus the usual definition of Euclidean distance error for the random vectors cannot be applied. To solve this problem, a distance named Optimal Sub-pattern Assignment (OSPA) is given in  \cite{Ref4}. OSPA is widely used in the performance analysis of algorithms (e.g. \cite{Ref5} and \cite{Ref6}), in the framework of the finite set statistics.

Based on the OSPA, a mean square error (MSE) between the state set
and estimation set is defined in this paper. We want to find a limit
of this MSE. When the state, measurement, and estimation are all
random vectors, the limit for MSE is called Posterior Cramer-Rao
bounds (PCRLB) \cite{Ref1}. In the framework of random vector
statistics, ${{\bf{z}}_k}$ is the measurement vector depends on the
state vector $ {{\bf{x}}_k} $ at time step $k$. The estimation
vector ${{\bf{\hat x}}_k}$ is based on the information gotten from
the measurement before time step $k+1$: ${{\bf{z}}_1},{{\bf{z}}_2},
\cdots ,{{\bf{z}}_k}$. Correspondingly, in the framework of the
finite set, the estimation set ${\hat X_k}$ is a function of all
measurement sets:
\begin{equation}\label{eq-1}
{Z_1},{Z_2}, \cdots ,{Z_k}.
\end{equation}
Therefore, the definition of MSE between the state set and estimation set relates to the serial measurement sets as in (\ref{eq-1}).

When the bound of the MSE is deduced, the PCRLB is also used. The
PCRLB in \cite{Ref7} is a fundamental contribution for the
development of the PCRLB. As the developments of the PCRLB in
\cite{Ref7}, in the case of ${P_d} < 1$ and ${P_{FA}} = 0$,
Information Reduction Factor (IRF) PCRLB \cite{Ref8} and enumeration
(ENUM) PCRLB \cite{Ref9} can be applied by considering the effect of
uncertainty in the measurement origin. Recently, these bounds are
further tightened in \cite{Ref10} and \cite{Ref11} respectively in
clutter environment. Comparing to other PCRLBs, the ENUM PCRLB is
the most accurate and the true bound for the case of ${P_d} < 1$ and
${P_{fa}} = 0$ \cite{Ref12}.

There is another error bound based on OSPA given in \cite{Ref13}
recently. This bound has a great influence on the derivation of the
error bound in this paper. However, the bound in \cite{Ref13} models
the state as a random set $X$, which is not a function of time step
$k$. In the other words, the bound in \cite{Ref13} is not recursive.
Obviously, the meaning of a non-recursive bound is limited to the
tracking system.

In this paper, for dynamic estimation and filtering problem, the state set $X_k$ is a Markov process. In order to discuss the appearance and disappearance of the targets, the $X_k$ may be $\left\{ {{{\bf{x}}_k}} \right\}$ or empty according to the probability. At the time step k, the measurement set is:
\begin{equation}\label{eq-2}
{Z_k} = \Theta \left( {{X_k}} \right).
\end{equation}
Moreover,  (\ref{eq-2}) is modeled in the case of ${P_d} < 1$, whose influence is significant to the calculation of error bounds.

In addition, part of this result in this paper has been reported in \cite{Ref15}, where only linear filtering case was presented and the discussion of the results was absent.

Section \ref{s2} revises the traditional PCRLB and the basic
knowledge of random set statistics. A new concept of mean square
error (MSE) $\sigma _k^2$ between $X_k$ and ${\hat X_k}$ is defined
in the section \ref{s3}. In the section \ref{s4}, a recursive form
of this error bound is derived. This bound is discussed in section
\ref{s5}. We present two numerical examples in the section \ref{s6}.
Proofs of the propositions are in the Section \ref{s7}. Conclusions
are drawn in the Section \ref{s8}.

\section{Background}\label{s2}

\subsection{Recursive Form of the PCRLB}
For a random vectors filtering problem, the state dynamic equation is given by:
\begin{equation}\label{eq-3}
{{\bf{x}}_{k + 1}} = {{\bf{f}}_k}\left( {{{\bf{x}}_k},{{\bf{w}}_k}} \right).
\end{equation}
where ${{\bf{f}}_k}$ is the state transition function, and ${{{\bf{w}}_k}}$ is a zero-mean white Gaussian process noise, with covariance matrix ${{\bf{Q}}_k}$.

When the target is detected, the measurement equation is given by:
\begin{equation}\label{eq-4}
{{\bf{z}}_k} = {{\bf{h}}_k}\left( {{{\bf{x}}_k},{{\bf{v}}_k}} \right).
\end{equation}
where ${{\bf{h}}_k}$ is the observation function, and ${{{\bf{v}}_k}}$ is a zero-mean white Gaussian noise, with covariance matrix ${{\bf{R}}_k}$.

Let ${{\bf{\hat x}}_k}$ be an unbiased state estimator based on the sequence of measurements $\left\{ {{{\bf{z}}_1}, \cdots ,{{\bf{z}}_k}} \right\}$.  The covariance of this estimator has a lower bound expressed as follows\cite{Ref1}:
\begin{equation}\label{eq-5}
E\left[ {\left( {{{{\bf{\hat x}}}_k} - {{\bf{x}}_k}} \right){{\left( {{{{\bf{\hat x}}}_k} - {{\bf{x}}_k}} \right)}^T}} \right] \ge {\bf{J}}_k^{ - 1}.
\end{equation}
where ${{\bf{J}}_k}$ is referred to as the Fisher information matrix (FIM), and  the ${{\bf{P}}_k} = {\bf{J}}_k^{ - 1}$ is the PCRLB.

As in \cite{Ref9}, when the target is detected, the recursive formula of FIM is as follow:
\begin{equation}\label{eq-6}
\begin{array}{l}
 {{\bf{J}}_{k + 1}} = {\bf{Q}}_k^{ - 1} + E\left\{ {{\bf{H}}_{k + 1}^T{\bf{R}}_{k + 1}^{ - 1}{\bf{H}}_{k + 1}^{}} \right\} \\
 \begin{array}{*{20}{c}}
   {} & {} & {}  \\
\end{array} - {\bf{Q}}_k^{ - 1}E\left\{ {{{\bf{F}}_k}} \right\}{\left[ {{{\bf{J}}_k} + E\left\{ {{\bf{F}}_k^T{\bf{Q}}_k^{ - 1}{{\bf{F}}_k}} \right\}} \right]^{ - 1}}E\left\{ {{{\bf{F}}_k}} \right\}{\bf{Q}}_k^{ - 1} \\
 \end{array}
\end{equation}
where the matrices ${{\bf{F}}_k}$ and ${{\bf{H}}_k}$ are respectively the Jacobians of  nonlinear functions ${{\bf{f}}_k}$ and ${{\bf{h}}_k}$:

\begin{equation}\label{eq-6a}
\begin{array}{l}
 {{\bf{F}}_k} = {\left[ {{\nabla _{{{\bf{x}}_k}}}{{\left[ {{{\bf{f}}_k}\left( {{{\bf{x}}_k}} \right)} \right]}^T}} \right]^T} \\
 {{\bf{H}}_k} = {\left[ {{\nabla _{{{\bf{x}}_k}}}{{\left[ {{{\bf{h}}_k}\left( {{{\bf{x}}_k}} \right)} \right]}^T}} \right]^T} \\
 \end{array}
\end{equation}

Also as in \cite{Ref9}, if the target is missed, or the FIM is for predictive, the recursive formula of  FIM reads
\begin{equation}\label{eq-7}
\begin{array}{l}
 {{\bf{J}}_{k + 1}} = {\bf{Q}}_k^{ - 1} \\
 \begin{array}{*{20}{c}}
   {} & {} & {}  \\
\end{array} - {\bf{Q}}_k^{ - 1}E\left\{ {{{\bf{F}}_k}} \right\}{\left[ {{{\bf{J}}_k} + E\left\{ {{\bf{F}}_k^T{\bf{Q}}_k^{ - 1}{{\bf{F}}_k}} \right\}} \right]^{ - 1}}E\left\{ {{{\bf{F}}_k}} \right\}{\bf{Q}}_k^{ - 1} \\
 \end{array}
\end{equation}

In a word, whether or not ${{\bf{z}}_k}$ ever exists, the FIM at $k+1$ can be calculated by the dynamic equation, measurement equation and the FIM at $k$.


\subsection{Random Finite Set}

Random finite set (RFS) is a random variable which takes value as finite set \cite{Ref13}. The element of this set is unordered random variable and the number of the elements is random and finite. Finite set statistics (FISST) is developed by Mahler \cite{Ref3} and widely considered an effective tool for the multi-target tracking system. In the perspective of modeling the tracking system, two types of RFS are often used: Poisson and Bernoulli RFS. Based on the model of Poisson RFS, a filter named Probability Hypothesis Density (PHD) filter \cite{Ref14} are applied in several fields \cite{Ref16}, \cite{Ref17}. However, PHD is a first-order statistical moment of the multi-target posterior \cite{Ref14}, and Poisson RFS is apt to model the multi-target tracking system. Therefore, Poisson RFS model does not suit to the single target appearance and disappearance problem in this paper.
The filter derived from Bernoulli RFS attracts substantial interest and is used widely recently \cite{Ref14}, \cite{Ref18}. As in \cite{Ref14}, here a Bernoulli RFS on a space $S$ is defined by two parameters $r$ and ${p\left(  \bullet  \right)}$:
\begin{equation}\label{eq-8}
f\left( X \right) = \left\{ {\begin{array}{*{20}{c}}
   {1 - b,} & {X = \emptyset ;}  \\
   {bp\left( {\bf{x}} \right),} & {X = \left\{ {\bf{x}} \right\};}  \\
   {0,} & {otherwise.}  \\
\end{array}} \right.
\end{equation}
where the $f\left( X \right)$ is the density of the RFS $X$ on the space of finite sets.

For the function $g$ taking value on the set $X$, the set integral of this function is \cite{Ref3}:
\begin{equation}\label{eq-9}
\int_S {g\left( X \right)\delta X}  \buildrel \Delta \over = g\left( \emptyset  \right) + \sum\limits_{n = 1}^\infty  {\frac{1}{{n!}}\int_{{S^n}} {g\left( {\left\{ {{{\bf{x}}_1}, \cdots ,{{\bf{x}}_n}} \right\}} \right)} } d{{\bf{x}}_1}, \cdots ,d{{\bf{x}}_n}
\end{equation}

The expectation of the function $g$ on a RFS of density $f$ is
\begin{equation}\label{eq-10}
E\left[ h \right] = \int_S {h\left( X \right)f\left( X \right)\delta X}
\end{equation}

If the state of the target is $X$, the estimation is ${\hat X\left( Z \right)}$, where $Z$ is the measurement of the target.
The distance defined between two sets $X$ and ${\hat X\left( Z \right)}$ is as follow \cite{Ref13}:

\begin{equation}\label{eq-11}
{\bf{e}}\left( {X = \left\{ {\bf{x}} \right\}, \hat X\left( Z \right) = \left\{ {{\bf{\hat x}}} \right\}} \right){\rm{ = }}{\bf{x}}{\rm{ - }}{\bf{\hat x}};
\end{equation}
\begin{equation}\label{eq-12}
e\left( {X = \emptyset ,\hat X\left( Z \right) = \left\{ {{\bf{x'}}}
\right\}} \right) \buildrel \Delta \over = {{\bf{e}}_0},for\mathop
{}\nolimits^{} any\mathop {}\nolimits^{} {\bf{x'}};
\end{equation}
\begin{equation}\label{eq-13}
e\left( {X = \left\{ {\bf{x}} \right\},\hat X\left( Z \right) =
\emptyset } \right) \buildrel \Delta \over = {{\bf{e}}_1},for\mathop
{}\nolimits^{} any\mathop {}\nolimits^{} {\bf{x}};
\end{equation}
\begin{equation}\label{eq-14}
{\bf{e}}\left( {X = \emptyset , \hat X\left( Z \right) = \emptyset } \right) = {\bf{0}}.
\end{equation}

Since the number of element in the set may be zero or one, the
difference between $X$ and ${\hat X\left( Z \right)}$ is defined in
(\ref{eq-12}) and (\ref{eq-13}), when the numbers of element of this
two sets are different. For one thing, if there is no target in
reality, we still estimate there is one target, the error is
${{\bf{e}}_0}$. Or perhaps, there is one target, but we estimate
there is no one, such error is ${{\bf{e}}_1}$. In a word,
${{\bf{e}}_0}$ and ${{\bf{e}}_1}$ indicate the mismatches of
cardinality.

According to the definition of the error, the mean square error between $X$ and ${\hat X\left( Z \right)}$ is given as in \cite{Ref13}:

\begin{equation}\label{eq-15}
\begin{array}{l}
 \Sigma  = E\left[ {{\bf{e}}\left( {X,\hat X\left( Z \right)} \right){\bf{e}}{{\left( {X,\hat X\left( Z \right)} \right)}^T}} \right] \\
  = \int {\int {{\bf{e}}\left( {X,\hat X\left( Z \right)} \right){\bf{e}}{{\left( {X,\hat X\left( Z \right)} \right)}^T}p\left( {X,Z} \right)} } \delta X\delta Z, \\
 \end{array}
\end{equation}
where ${p\left( {X,Z} \right)}$ is the joint density of the state set $X$ and measurement set $Z$.

\section{Necessary Definition}\label{s3}

\subsection{Observation-sets Sequence}
In the framework of random vector statistics, at time step $k$, the
estimation is a function of all measurements from time step $1$ to
$k$, and contain all information of such measurements. These
measurements are ${{\bf{z}}_1},{{\bf{z}}_2}, \cdots ,{{\bf{z}}_k}$,
and thus the estimation is ${{\bf{\hat x}}_k}\left( {\left\{
{{{\bf{z}}_1}, \cdots ,{{\bf{z}}_k}} \right\}} \right)$.
Correspondingly, in the framework of random set, firstly, we should
determine how to state the measurement sets from time step $1$ to
$k$, which is defined as observation-sets sequence ${\Theta
_{k,n}}$.

At time-step $k$, the possible time-sequence of observation-sets is given as follows:
\begin{equation}\label{eq-16}
{\Theta _{k,n}} = \left\{ {{Z_{1,n}},{Z_{2,n}}, \cdots ,{Z_{k,n}}} \right\},
\end{equation}
where ${Z_{k,n}}$ denotes the measurement is empty or not for  sequence number $n$, at time-step $k$, and thus, $n = 1,2, \cdots ,{2^k}$.

In order to not only simplify the form of the bound deduced in this paper, but also indicate the influence of whether the observation is empty or not, the arrangement of the elements in ${\Theta _{k,n}}$ is not random, but follows the rule:

When $k=1$
\begin{equation}\label{eq-17}
{\Theta _{1,1}} = \emptyset ,{\Theta _{1,2}} = \left\{ {{{\bf{z}}_1}} \right\}.
\end{equation}

When $k=2$
\begin{equation}\label{eq-18}
\begin{array}{l}
 {\Theta _{2,1}} = \left\{ {\emptyset ,\emptyset } \right\},{\Theta _{2,2}} = \left\{ {\left\{ {{{\bf{z}}_1}} \right\},\emptyset } \right\}, \\
 {\Theta _{2,3}} = \left\{ {\emptyset ,\left\{ {{{\bf{z}}_2}} \right\}} \right\},{\Theta _{2,4}} = \left\{ {\left\{ {{{\bf{z}}_1}} \right\},\left\{ {{{\bf{z}}_2}} \right\}} \right\}. \\
 \end{array}
\end{equation}

When $k>2$
\begin{equation}\label{eq-19}
\begin{array}{l}
 {\Theta _{k + 1,n}} \\
  = \left\{ {\begin{array}{*{20}{c}}
   {\left\{ {{\Theta _{k,n}},\emptyset } \right\},} & {1 \le n \le {2^k}}  \\
   {\left\{ {{\Theta _{k,n - {2^k}}},\left\{ {{z_k}} \right\}} \right\},} & {{2^k} + 1 \le n \le {2^{k + 1}}}  \\
\end{array}} \right. \\
  = \left\{ {\begin{array}{*{20}{c}}
   {\left\{ {{\Theta _{k - 1,n}},\emptyset ,\emptyset } \right\},} & {1 \le n \le {2^{k - 1}}}  \\
   {\left\{ {{\Theta _{k - 1,n - {2^{k - 1}}}},\left\{ {{z_k}} \right\},\emptyset } \right\},} & {{2^{k - 1}} + 1 \le n \le {2^k}}  \\
   {\left\{ {{\Theta _{k - 1,n - {2^k}}},\emptyset ,\left\{ {{z_k}} \right\}} \right\},} & {{2^k} + 1 \le n \le {2^k} + {2^{k - 1}}}  \\
   {\left\{ \begin{array}{l}
 {\Theta _{k - 1,n - {2^k} - {2^{k - 1}}}}, \\
 \left\{ {{z_k}} \right\},\left\{ {{z_{k + 1}}} \right\} \\
 \end{array} \right\},} & {{2^k} + {2^{k - 1}} + 1 \le n \le {2^{k + 1}}}  \\
\end{array}} \right. \\
 \end{array}
\end{equation}

(\ref{eq-19}) shows that the elements in ${\Theta_{k+1,n}}$ can be divided into four parts with equal number of elements, according to the measurements at time step $k$ and $k+1$. If the dividing is just on the basis of the measurement at time step $k+1$, there are two parts. The first one is in the situation that the measurement set is empty ${Z_{k + 1,n}} = \emptyset $, where the sequence number $n$ is in the scope $1 \le n \le {2^k}$. This situation appears when there is no target or the target is missed. The second part is in the condition that ${Z_{k + 1,n}} \ne \emptyset $, where  $n$ is in the range ${2^k} + 1 \le n \le {2^{k + 1}}$. This condition results from that there is a target and it has been observed.

It is notable that, at time step $k+1$, the ${\Theta _{k,n - {2^k}}}$ with ${2^k} + 1 \le n \le {2^{k + 1}}$ and the ${\Theta _{k,n}}$ with $1 \le n \le {2^k}$ share the same observation-sets sequence, which is all possible observation-sets sequence at time step $k$.


\subsection{Error Bounds Defined based on Observation-sets Sequence}
For certain time-sequence of observation-sets ${\Theta _{k,n}}$, the estimation can be written into a particular form. Then, the estimated error is defined as following:
\begin{equation}\label{eq-20}
{\Sigma _{k,n}} = \int { \cdots \int {{{\bf{C}}_{k,n}}p({X_k},{\Theta _{k,n}})\delta {X_k}\delta {Z_{1,n}} \cdots \delta {Z_{k,n}}} }  \ge {{\bf{P}}_{k,n}}
\end{equation}
where
\begin{equation}\label{eq-21}
{{\bf{C}}_{k,n}} \buildrel \Delta \over = {\bf{e}}\left( {{X_k},{{\hat X}_k}({\Theta _{k,n}})} \right){\bf{e}}{\left( {{X_k},{{\hat X}_k}({\Theta _{k,n}})} \right)^T}
\end{equation}

At time step $k$, ${{\bf{P}}_{k,n}}$ is the error bound for the particular observation-sets ${\Theta _{k,n}}$.

Because ${\Theta _{k,n}}$ can cover all possible conditions of observation when the number of sequence $n$ takes value from $1$ to ${2^{k}}$, the total error between ${X_k}$ and its estimation ${\hat X_k}({Z_1} \cdots {Z_k})$ is ${\Sigma _{k,n}}$:
\begin{equation}\label{eq-22}
\begin{array}{l}
 {\Sigma _k} = \int { \cdots \int \begin{array}{l}
 {\bf{e}}\left( {{X_k},{{\hat X}_k}({Z_1} \cdots {Z_k})} \right){\bf{e}}{\left( {{X_k},{{\hat X}_k}({Z_1} \cdots {Z_k})} \right)^T}* \\
 p({X_k},{Z_1}, \cdots ,{Z_k})\delta {X_k}\delta {Z_1} \cdots \delta {Z_k} \\
 \end{array} }  \\
  = \sum\limits_{n = 1}^{{2^k}} {\int { \cdots \int \begin{array}{l}
 {\bf{e}}\left( {{X_k},{{\hat X}_k}({\Theta _{k,n}})} \right){\bf{e}}{\left( {{X_k},{{\hat X}_k}({\Theta _{k,n}})} \right)^T}* \\
 p({X_k},{\Theta _{k,n}})\delta {X_k}\delta {Z_{1,n}} \cdots \delta {Z_{k,n}} \\
 \end{array} } }  \\
  = \sum\limits_{n = 1}^{{2^k}} {{\Sigma _{k,n}}}.  \\
 \end{array}
\end{equation}

Therefore, the error bound of such total error is as follow:
\begin{equation}\label{eq-23}
{\Sigma _k} = \sum\limits_{n = 1}^{{2^k}} {{\Sigma _{k,n}}}  \ge {{\bf{P}}_k} = \sum\limits_{n = 1}^{{2^k}} {{{\bf{P}}_{k,n}}}.
\end{equation}

Hence, in the rest of paper, what we deduce is the recursive form of
${{{\bf{P}}_{k,n}}}$. According to (\ref{eq-23}), the total bound
${{\bf{P}}_k}$ can be obtained by the sum of ${{{\bf{P}}_{k,n}}}$.

\section{Recursive Form of the Bound}\label{s4}

\subsection{Random Finite Set Models}
Since the target in either 'present' or 'absent' state, the state of the target is modeled by Bernoulli RFS mentioned in section \ref{s2}.

For the dynamical model, the Markov transition density is defined by:
\begin{equation}\label{eq-24}
f({X_{k + 1}}\left| {{X_k} = \left\{ {{{\bf{x}}_k}} \right\}} \right.) = \left\{ {\begin{array}{*{20}{c}}
   {r\psi ({{\bf{x}}_{k + 1}}\left| {{{\bf{x}}_k}} \right.),} & {{X_{k + 1}} = \left\{ {{{\bf{x}}_{k + 1}}} \right\}}  \\
   {1 - r,{\rm{ }}} & {{X_{k + 1}} = \emptyset }  \\
\end{array}} \right.
\end{equation}
\begin{equation}\label{eq-25}
f({X_{k + 1}}\left| {{X_k} = \emptyset } \right.) = \left\{ {\begin{array}{*{20}{c}}
   {\left( {1 - r} \right){p_0}\left( {{{\bf{x}}_{k + 1}}} \right),} & {{X_{k + 1}} = \left\{ {{{\bf{x}}_{k + 1}}} \right\}}  \\
   {r,} & {{X_{k + 1}} = \emptyset }  \\
\end{array}\begin{array}{*{20}{c}}
   {\rm{ }}  \\
   {\rm{ }}  \\
\end{array}} \right.
\end{equation}
where $r \in \left[ {0,1} \right]$ represents the probability of the
state of the target at time step $k + 1$ survives from the state at
time-step $k$ or remains empty. It means, conditional upon ${X_k} =
\left\{ {{{\bf{x}}_k}} \right\}$, this target disappear with the
probability $1-r$. If there is no target at time-step , a new target
would bear with the probability $1-r$, and the initial density is
${p_0}\left( {{{\bf{x}}_{k + 1}}} \right)$. Therefore, $r$ is named
as the maintenance probability. $\psi ({{\bf{x}}_{k + 1}}\left|
{{{\bf{x}}_k}} \right.)$ is the probability density of a transition
from ${{\bf{x}}_{k}}$ state to ${{\bf{x}}_{k+1 }}$.

The prior probability function of the state set is also Bernoulli RFS:
\begin{equation}\label{eq-26}
f\left( {{X_0}} \right) = \left\{ {\begin{array}{*{20}{c}}
   {b{p_0}\left( {{{\bf{x}}_0}} \right),{X_0} = \left\{ {{{\bf{x}}_0}} \right\}}  \\
   {1 - b,{X_0} = \emptyset }  \\
\end{array}} \right.
\end{equation}
where $b \in \left[ {0,1} \right]$ representing the probability of the target existing initially. $b$ is named as the initial probability.

The probability of detection is ${P_d} < 1$. The measurement model is:
\begin{equation}\label{eq-27}
g({Z_k}\left| {{X_k} = \left\{ {{{\bf{x}}_k}} \right\}} \right.) = \left\{ {\begin{array}{*{20}{c}}
   {{P_d}\xi ({{\bf{z}}_k}\left| {{{\bf{x}}_k}} \right.),{\rm{ }}{Z_k} = \left\{ {{{\bf{z}}_k}} \right\}}  \\
   {1 - {P_d},{\rm{ }}{Z_k} = \emptyset }  \\
\end{array}} \right.
\end{equation}
\begin{equation}\label{eq-28}
g({Z_k}\left| {{X_k} = \emptyset } \right.) = \left\{ {\begin{array}{*{20}{c}}
   {0,{\rm{ }}{Z_k} = \left\{ {{{\bf{z}}_k}} \right\}}  \\
   {1,{\rm{ }}{Z_k} = \emptyset }  \\
\end{array}} \right.
\end{equation}
where $\xi ({{\bf{z}}_k}\left| {{{\bf{x}}_k}} \right.)$ is the measurement likelihood when the target is existing and detected. (\ref{eq-27}) indicates there is some uncertainty in detection, and (\ref{eq-28}) means there is no false observation.

\subsection{Derivation of the Bound}

\subsubsection{Derivation of ${{\bf{P}}_k}$}
As discussion in section \ref{s3}, in order to calculate the recursive form of ${{\bf{P}}_k}$, we should get the recursive ${{\bf{P}}_{k,n}}$.

\emph{Proposition I}: When time-step $k \ge 0$, $n = 1, \cdots
,{2^{k + 1}}$ , the bounds sequence ${{\bf{P}}_{k + 1,n}}$ obeys the
recursion:

\begin{equation}\label{eq-29}
{{\bf{P}}_{k + 1,n}} = \left\{ {\begin{array}{*{20}{c}}
   {{\bf{\tilde P}}_{k + 1,n}^{},\begin{array}{*{20}{c}}
   {} & {}  \\
\end{array}1 \le n \le {2^k}}  \\
   \begin{array}{l}
 {\left[ {{{\bf{J}}_{k + 1,n}}} \right]^{ - 1}}*\Pr \left( {{\Theta _{k,n - {2^k}}},{Z_{k + 1}} \ne \emptyset } \right){\rm{, }} \\
 \begin{array}{*{20}{c}}
   {} & {} & {} & {}  \\
\end{array}\begin{array}{*{20}{c}}
   {} & {} & {} & {}  \\
\end{array}{2^k} + 1 \le n \le {2^{k + 1}} \\
 \end{array}  \\
\end{array}} \right.
\end{equation}

\begin{equation}\label{eq-30}
{\bf{\tilde P}}_{k + 1,n}^{} = \left\{ {\begin{array}{*{20}{c}}
   {{\bf{P}}_{k + 1,n}^*,} & {trace({\bf{P}}_{k + 1,n}^*) < trace{\bf{(P}}_{k + 1,n}^{**})}  \\
   {{\bf{P}}_{k + 1,n}^{**},} & {otherwise}  \\
\end{array}} \right.
\end{equation}
where
\begin{equation}\label{eq-31}
{\bf{P}}_{k + 1,n}^* = {\bf{e}}_1^{}{\bf{e}}_1^T\left( {\Pr \left( {{\Theta _{k,n}},{Z_{k + 1}} = \emptyset } \right) - {\rho _{k + 1,n}}} \right)
\end{equation}

\begin{equation}\label{eq-32}
{\bf{P}}_{k + 1,n}^{**} = {\bf{e}}_0^{}{\bf{e}}_0^T{\rho _{k + 1,n}} + {\left[ {{{\bf{J}}_{k + 1,n}}} \right]^{ - 1}}*\Pr \left( {{\Theta _{k,n}},{Z_{k + 1}} = \emptyset } \right)
\end{equation}

\begin{equation}\label{eq-33}
\begin{array}{l}
 {\rho _{k + 1,n}} \\
  \buildrel \Delta \over = \int { \cdots \int {p({X_{k + 1}} = \emptyset ,{\Theta _{k,n}},{Z_{k + 1,n}} = \emptyset )\delta {Z_{1,n}} \cdots \delta {Z_{k,n}}} }  \\
 \end{array}
\end{equation}

${\rho _{k + 1,n}}$ indicates the probability of the state set of the target is empty at time step $k+1$, when all the measurements from time step $1$ to $k+1$ are all known.

The proof of this proposition is in the section \ref{s7}.

From proposition I , we can see that the problem of recursion of ${{\bf{P}}_{k,n}}$ reduces to the recursion of ${{\bf{J}}_{k + 1,n}}$, ${\rho _{k + 1,n}}$ and $\Pr \left( {{\Theta _{k + 1,n}}} \right)$. Then we deduce how to obtain the recurrent formulas of all these three factors.
First, based on (\ref{eq-6}) and (\ref{eq-7}), for a particular ${\Theta _{k,n}}$ defined in (\ref{eq-19}), the FIM ${{\bf{J}}_{k + 1,n}}$ can be calculated from ${{\bf{J}}_{k,n}}$ when $k \ge 0$:

\begin{equation}\label{eq-35}
\begin{array}{l}
 {{\bf{J}}_{k + 1,n}} =  \\
 \left\{ {\begin{array}{*{20}{c}}
   {{\bf{Q}}_k^{ - 1} - {\bf{Q}}_k^{ - 1}E\left\{ {{{\bf{F}}_k}} \right\}{{\left[ {{{\bf{J}}_{k,n}} + E\left\{ {{\bf{F}}_k^T{\bf{Q}}_k^{ - 1}{{\bf{F}}_k}} \right\}} \right]}^{ - 1}}E\left\{ {{{\bf{F}}_k}} \right\}{\bf{Q}}_k^{ - 1}{\rm{,}}}  \\
   {\begin{array}{*{20}{c}}
   {} & {} & {} & {}  \\
\end{array}\begin{array}{*{20}{c}}
   {} & {} & {} & {}  \\
\end{array}\begin{array}{*{20}{c}}
   {} & {} & {} & {}  \\
\end{array}\begin{array}{*{20}{c}}
   {} & {}  \\
\end{array}1 \le n \le {2^k}}  \\
   \begin{array}{l}
 {\bf{Q}}_k^{ - 1} + E\left\{ {{\bf{H}}_{k + 1}^T{\bf{R}}_{k + 1}^{ - 1}{\bf{H}}_{k + 1}^{}} \right\} \\
  - {\bf{Q}}_k^{ - 1}E\left\{ {{{\bf{F}}_k}} \right\}{\left[ {{{\bf{J}}_{k,n - {2^k}}} + E\left\{ {{\bf{F}}_k^T{\bf{Q}}_k^{ - 1}{{\bf{F}}_k}} \right\}} \right]^{ - 1}}E\left\{ {{{\bf{F}}_k}} \right\}{\bf{Q}}_k^{ - 1}, \\
 \end{array}  \\
   {\begin{array}{*{20}{c}}
   {} & {} & {} & {}  \\
\end{array}\begin{array}{*{20}{c}}
   {} & {} & {} & {}  \\
\end{array}\begin{array}{*{20}{c}}
   {} & {} & {} & {}  \\
\end{array}{2^k} + 1 \le n \le {2^{k + 1}}}  \\
\end{array}} \right. \\
 \end{array}
\end{equation}

If there is no process noise ${{{\bf{w}}_k}}$, then (\ref{eq-35}) reduces to:

\begin{equation}\label{eq-36}
{{\bf{J}}_{k + 1,n}} = \left\{ {\begin{array}{*{20}{c}}
   {{{\left[ {{\bf{F}}_k^{ - 1}} \right]}^T}{{\bf{J}}_{k,n}}{\bf{F}}_k^{ - 1},1 \le n \le {2^k}}  \\
   \begin{array}{l}
 {\left[ {{\bf{F}}_k^{ - 1}} \right]^T}{{\bf{J}}_{k,n - {2^k}}}{\bf{F}}_k^{ - 1} +  \\
 {\bf{H}}_{k + 1}^T{\bf{R}}_{k + 1}^{ - 1}{\bf{H}}_{k + 1}^{},{2^k} + 1 \le n \le {2^{k + 1}} \\
 \end{array}  \\
\end{array}} \right.
\end{equation}

The initial FIM is calculated from the prior probability function ${p_0}\left( {{{\bf{x}}_0}} \right)$:
\begin{equation}\label{eq-37}
{{\bf{J}}_{0,1}} = E\left\{ { - \Delta _{{{\bf{x}}_0}}^{{{\bf{x}}_0}}\log {p_0}\left( {{{\bf{x}}_0}} \right)} \right\}
\end{equation}

\subsubsection{Derivation of $\Pr \left( {{\Theta _{k + 1,n}}} \right)$}
Secondly, $\Pr \left( {{\Theta _{k + 1,n}}} \right)$ consists two parts: $\Pr \left( {{\Theta _{k,n}},{Z_{k + 1}} = \emptyset } \right)$ and $\Pr \left( {{\Theta _{k,n - {2^k}}},{Z_{k + 1}} \ne \emptyset } \right)$, according to the ${\Theta _{k+1,n}}$.
When $k=0$, the initial probability of the empty measurement is as follow:
\begin{equation}\label{eq-38}
\Pr \left( {{\Theta _{0,1}},{Z_1} = \emptyset } \right) = \Pr \left( {{\Theta _{1,1}}} \right) = 1 - b*{P_d}
\end{equation}
\begin{equation}\label{eq-39}
\Pr \left( {{\Theta _{0,1}},{Z_1} \ne \emptyset } \right) = \Pr \left( {{\Theta _{1,2}}} \right) = b*{P_d}
\end{equation}

When $k \ge 1$, the recursion of $\Pr \left( {{\Theta _{k + 1,n}}} \right)$ is given in Proposition II:

\emph{Proposition II}: When $k \ge 1$
\begin{equation}\label{eq-40}
\begin{array}{l}
 \Pr \left( {{\Theta _{k + 1,n}}} \right) \\
  = \left\{ {\begin{array}{*{20}{c}}
   \begin{array}{l}
 \Pr \left( {{\Theta _{k,n}}} \right)p\left( {{Z_{k + 1,n}} = \emptyset \left| {{\Theta _{k,n}}} \right.} \right), \\
 \begin{array}{*{20}{c}}
   {} & {} & {} & {}  \\
\end{array}\begin{array}{*{20}{c}}
   {} & {} & {} & {}  \\
\end{array}\begin{array}{*{20}{c}}
   {} & {}  \\
\end{array}1 \le n \le {2^{k - 1}} \\
 \end{array}  \\
   \begin{array}{l}
 \Pr \left( {{\Theta _{k,n}}} \right)\left( {1 - r{P_d}} \right), \\
 \begin{array}{*{20}{c}}
   {} & {} & {} & {}  \\
\end{array}\begin{array}{*{20}{c}}
   {} & {}  \\
\end{array}\begin{array}{*{20}{c}}
   {} & {}  \\
\end{array}{2^{k - 1}} + 1 \le n \le {2^k} \\
 \end{array}  \\
   \begin{array}{l}
 \Pr \left( {{\Theta _{k,n - {2^k}}}} \right)\left[ {1 - p\left( {{Z_{k + 1,n}} = \emptyset \left| {{\Theta _{k,n - {2^k}}}} \right.} \right)} \right], \\
 \begin{array}{*{20}{c}}
   {} & {} & {} & {}  \\
\end{array}\begin{array}{*{20}{c}}
   {} & {}  \\
\end{array}{2^k} + 1 \le n \le {2^k} + {2^{k - 1}} \\
 \end{array}  \\
   \begin{array}{l}
 \Pr \left( {{\Theta _{k,n - {2^k}}}} \right)r{P_d}, \\
 \begin{array}{*{20}{c}}
   {} & {} & {} & {}  \\
\end{array}\begin{array}{*{20}{c}}
   {} & {}  \\
\end{array}{2^k} + {2^{k - 1}} + 1 \le n \le {2^{k + 1}} \\
 \end{array}  \\
\end{array}} \right. \\
 \end{array}
\end{equation}

The proposition II is proved in the section \ref{s7}.

At time step $k+1$, for the number of queue $n$ in the two range
that ${2^{k - 1}} + 1 \le n \le {2^k}$ and ${2^k} + {2^{k - 1}} + 1
\le n \le {2^{k + 1}}$, the measurement ${Z_{k,n}} \ne \emptyset$.
It means that, the target exists at time step $k$, because it is
assumed that there is no clutter. Then, at time step k+1, the state
of the target can be written by the state transition model. Hence,
$p\left( {{Z_{k + 1,n}} \ne \emptyset \left| {{\Theta _{k,n}}}
\right.} \right)$ can be calculated easily. However, when the
measurement ${Z_{k,n}} = \emptyset$, which corresponds the two range
that $1 \le n \le {2^{k - 1}}$ and ${2^k} + 1 \le n \le {2^k} +
{2^{k - 1}}$, it is uncertain the reason is the state set is empty
or there is a miss detection. Therefore, it is hard to determine
$p\left( {{Z_{k + 1,n}} = \emptyset \left| {{\Theta _{k,n}}}
\right.} \right)$ or $p\left( {{Z_{k + 1,n}} \ne \emptyset \left|
{{\Theta _{k,n}}} \right.} \right)$, when ${Z_{k,n}} = \emptyset$.

The recursion of $p\left( {{Z_{k + 1,n}} = \emptyset \left| {{\Theta _{k,n}}} \right.} \right)$ is given in the proposition IV, for the recursive form of ${\rho _{k + 1,n}}$ is presented in the proposition III, which is also related to $p\left( {{Z_{k + 1,n}} = \emptyset \left| {{\Theta _{k,n}}} \right.} \right)$.

\subsubsection{Derivation of ${\rho _{k + 1,n}}$}

${\rho _{k + 1,n}}$ is defined in (\ref{eq-33}). When $k=0$, the initial ${\rho _{1,1}}$ is calculated as:
\begin{equation}\label{eq-41}
{\rho _{1,1}} = 1 - b
\end{equation}

When $k \ge 1$, ${\rho _{k + 1,n}}$ is calculated as the proposition III.

\emph{Proposition III}: When $k \ge 1$
\begin{equation}\label{eq-42}
\begin{array}{l}
 {\rho _{k + 1,n}} =  \\
 \Pr ({\Theta _{k,n}})\int { \cdots \int {\frac{{p\left( {{Z_{k + 1,n}} = \emptyset \left| {{\Theta _{k,n}}} \right.} \right) - \left( {1 - {P_d}} \right)}}{{{P_d}}}} } \delta {Z_{1,n}} \cdots \delta {Z_{k,n}}, \\
 \begin{array}{*{20}{c}}
   {} & {} & {} & {}  \\
\end{array}\begin{array}{*{20}{c}}
   {} & {} & {} & {}  \\
\end{array}\begin{array}{*{20}{c}}
   {} & {} & {} & {}  \\
\end{array}\begin{array}{*{20}{c}}
   {} & {} & {} & {}  \\
\end{array}1 \le n \le {2^k} \\
 \end{array}
\end{equation}

The proposition III is testified in the section \ref{s7}.

When the measurements from time step $1$ to $k+1$ have all obtained, ${\rho _{k + 1,n}}$ shows the probability of the state set is empty at time step $k+1$,. Hence, it is obvious that ${\rho _{k + 1,n}}$ is a function of $p\left( {{Z_{k + 1,n}} = \emptyset \left| {{\Theta _{k,n}}} \right.} \right)$.

\subsubsection{Derivation of $p\left( {{Z_{k + 1,n}} = \emptyset \left| {{\Theta _{k,n}}} \right.} \right)$}
From the proposition II and proposition III, we can see that both $\Pr \left( {{\Theta _{k + 1,n}}} \right)$ and ${\rho _{k + 1,n}}$ are functions of $p\left( {{Z_{k + 1,n}} = \emptyset \left| {{\Theta _{k,n}}} \right.} \right)$ when $1 \le n \le {2^k}$.
Therefore, the key is how to get the recursion of $p\left( {{Z_{k + 1,n}} = \emptyset \left| {{\Theta _{k,n}}} \right.} \right)$.

\emph{Proposition IV}: When $k \ge 1$
\begin{equation}\label{eq-43}
\begin{array}{l}
 p\left( {{Z_{k + 1,n}} = \emptyset \left| {{\Theta _{k,n}}} \right.} \right) \\
  = \left\{ {\begin{array}{*{20}{c}}
   \begin{array}{l}
 1 - r{P_d} + {P_d}\left( {2r - 1} \right)* \\
 \frac{{p\left( {{Z_{k,n}} = \emptyset \left| {{\Theta _{k - 1,n}}} \right.} \right) - \left( {1 - {P_d}} \right)}}{{{P_d}p\left( {{Z_{k,n}} = \emptyset \left| {{\Theta _{k - 1,n}}} \right.} \right)}}, \\
 \end{array} & {{\rm{ }}1 \le n \le {2^{k - 1}}}  \\
   {1 - r{P_d},} & {{2^{k - 1}} \le n \le {2^k}}  \\
\end{array}} \right.{\rm{   }} \\
  \buildrel \Delta \over = \Gamma \left( {p\left( {{Z_{k,n}} = \emptyset \left| {{\Theta _{k - 1,n}}} \right.} \right)} \right) \\
 \end{array}
\end{equation}

The proposition IV is testified in the section \ref{s7}.

By taking (\ref{eq-43}) into the Proposition II and III, $\Pr \left(
{{\Theta _{k + 1,n}}} \right)$ and ${\rho _{k + 1,n}}$ can be
obtained from which at the time step $k$. Therefore, the
${{\bf{P}}_{k + 1,n}}$ can be calculated in a recurrent form. From
the (\ref{eq-23}), the total error bound ${{\bf{P}}_k}$ can be
obtained.

\section{Discussion}\label{s5}
\subsection{The Meaning of This Bound}
If the state of target ${{\bf{x}}_k}$ is one-dimensional, the bounds calculated with empty measurements (\ref{eq-30}) turn to:

\begin{equation}\label{eq-43a}
{\bf{\tilde P}}_{k + 1,n}^{} = \left\{ {\begin{array}{*{20}{c}}
   {{\bf{P}}_{k + 1,n}^*,{\bf{P}}_{k + 1,n}^* < {\bf{P}}_{k + 1,n}^{**}}  \\
   {{\bf{P}}_{k + 1,n}^{**},{\bf{P}}_{k + 1,n}^* \ge {\bf{P}}_{k + 1,n}^{**}}  \\
\end{array}} \right.
\end{equation}

When ${\bf{\tilde P}}_{k + 1,n}^{} = {\bf{P}}_{k + 1,n}^*$, it means that

\begin{equation}\label{eq-43b}
\begin{array}{l}
 {\bf{e}}_0^{}{\bf{e}}_0^T{\rho _{k + 1,n}} + {\left[ {{{\bf{J}}_{k + 1,n}}} \right]^{ - 1}}*\Pr \left( {{\Theta _{k,n}},{Z_{k + 1}} = \emptyset } \right) >  \\
 {\bf{e}}_1^{}{\bf{e}}_1^T\left( {\Pr \left( {{\Theta _{k,n}},{Z_{k + 1}} = \emptyset } \right) - {\rho _{k + 1,n}}} \right) \\
 \end{array}
\end{equation}

Combining the definition of ${\rho _{k + 1,n}}$ in  (\ref{eq-33}), setting ${\bf{e}}_1^{} = {\bf{e}}_0^{}$, it is easy to derive that

\begin{equation}\label{eq-43c}
\begin{array}{l}
 \int { \cdots \int {f({X_{k + 1}} = \emptyset ,{\Theta _{k,n}},{Z_{k + 1,n}} = \emptyset )\delta {Z_{1,n}} \cdots \delta {Z_{k,n}}} }  >  \\
 \frac{{{\bf{e}}_1^{}{\bf{e}}_1^T - {{\left[ {{{\bf{J}}_{k + 1,n}}} \right]}^{ - 1}}}}{{{\bf{e}}_1^{}{\bf{e}}_1^T + {{\left[ {{{\bf{J}}_{k + 1,n}}} \right]}^{ - 1}}}}\int { \cdots \int \begin{array}{l}
 f({X_{k + 1}} \ne \emptyset ,{\Theta _{k,n}},{Z_{k + 1,n}} = \emptyset ) \\
 \delta {Z_{1,n}} \cdots \delta {Z_{k,n}} \\
 \end{array} }  \\
 \end{array}
\end{equation}

While the number of the scans of measurements increases, it should meet the condition that
\begin{equation}\label{eq-43d}
{\bf{e}}_0^{}{\bf{e}}_0^T = {\bf{e}}_1^{}{\bf{e}}_1^T \gg {\left[ {{{\bf{J}}_{k + 1,n}}} \right]^{ - 1}}
\end{equation}

As a result:
\begin{equation}\label{eq-43e}
\begin{array}{l}
 \int { \cdots \int {f({X_{k + 1}} = \emptyset ,{\Theta _{k,n}},{Z_{k + 1,n}} = \emptyset )\delta {Z_{1,n}} \cdots \delta {Z_{k,n}}} }  >  \\
 \int { \cdots \int {f({X_{k + 1}} \ne \emptyset ,{\Theta _{k,n}},{Z_{k + 1,n}} = \emptyset )\delta {Z_{1,n}} \cdots \delta {Z_{k,n}}} }  \\
 \end{array}
\end{equation}

It denotes that, when the probability of empty state set is more than which of not empty, the bound is in the form that ${\bf{\tilde P}}_{k + 1,n}^{} = {\bf{e}}_1^{}{\bf{e}}_1^T\left( {\Pr \left( {{\Theta _{k,n}},{Z_{k + 1}} = \emptyset } \right) - {\rho _{k + 1,n}}} \right)$. In the other word, there is ${\hat X_{k + 1}}({\Theta _{k,n}},{Z_{k + 1,n}} = \emptyset ) = \emptyset$, when the bound is attained.

On the other side, if (\ref{eq-43d}) is satisfied, and if
\begin{equation}\label{eq-43f}
\begin{array}{l}
 \int { \cdots \int {f({X_{k + 1}} \ne \emptyset ,{\Theta _{k,n}},{Z_{k + 1,n}} = \emptyset )\delta {Z_{1,n}} \cdots \delta {Z_{k,n}}} }  >  \\
 \int { \cdots \int {f({X_{k + 1}} = \emptyset ,{\Theta _{k,n}},{Z_{k + 1,n}} = \emptyset )\delta {Z_{1,n}} \cdots \delta {Z_{k,n}}} }  ,\\
 \end{array}
\end{equation}
the estimation is ${\hat X_{k + 1}}({\Theta _{k,n}},{Z_{k + 1,n}} = \emptyset ) \ne \emptyset $, and the bound is that ${\bf{\tilde P}}_{k + 1,n}^{} = {\bf{e}}_0^{}{\bf{e}}_0^T{\rho _{k + 1,n}} + {\left[ {{{\bf{J}}_{k + 1,n}}} \right]^{ - 1}}*\Pr \left( {{\Theta _{k,n}},{Z_{k + 1}} = \emptyset } \right)$. Then, this bound can be compared with the PCRLB.

\subsection{Comparison with Previous Results}
The enumeration PCRLB has been verified as the exact bound in the case of ${P_d} < 1$, both in a linear and a nonlinear case. It is the optimal lower bound for tracking in the framework of the finite vector statistics. Rewrite the PCRLB computed via enumeration in \cite{Ref9} and \cite{Ref12} as following:
\begin{equation}\label{eq-d1}
\begin{array}{l}
{P_{k + 1}}(ENUM) = \\
\sum\limits_{n = 1}^{{2^k}} {\left\{ \begin{array}{l} {\left[
\begin{array}{l}
{\bf{Q}}_k^{ - 1} - {\bf{Q}}_k^{ - 1}E\left\{ {{{\bf{F}}_k}} \right\} \times \\
{\left[ {{{\bf{J}}_{k,n}} + E\left\{ {{\bf{F}}_k^T{\bf{Q}}_k^{ -
1}{{\bf{F}}_k}} \right\}} \right]^{ - 1}}E\left\{ {{{\bf{F}}_k}}
\right\}{\bf{Q}}_k^{ - 1}
\end{array} \right]^{ - 1}}\\
 \times \Pr \left( {{\Theta _{k,n}},{Z_{k + 1}} = \emptyset } \right)
\end{array} \right\}}  + \\
\sum\limits_{n = {2^k} + 1}^{{2^{k + 1}}} {\left\{ \begin{array}{l}
{\left[ \begin{array}{l}
{\bf{Q}}_k^{ - 1} + E\left\{ {{\bf{H}}_{k + 1}^T{\bf{R}}_{k + 1}^{ - 1}{\bf{H}}_{k + 1}^{}} \right\}\\
 - {\bf{Q}}_k^{ - 1}E\left\{ {{{\bf{F}}_k}} \right\} \times \\
{\left[ {{{\bf{J}}_{k,n - {2^k}}} + E\left\{
{{\bf{F}}_k^T{\bf{Q}}_k^{ - 1}{{\bf{F}}_k}} \right\}} \right]^{ -
1}}E\left\{ {{{\bf{F}}_k}} \right\}{\bf{Q}}_k^{ - 1}
\end{array} \right]^{ - 1}}\\
 \times \Pr \left( {{\Theta _{k,n - {2^k}}},{Z_{k + 1}} \ne \emptyset } \right)
\end{array} \right\}}
\end{array}
\end{equation}

At the same time, if (\ref{eq-43d}) and (\ref{eq-43f}) are satisfied, our bound is given by:

\begin{equation}\label{eq-d2}
\begin{array}{l}
{P_{k + 1}}{\rm{RFS}} = \\
\sum\limits_{n = 1}^{{2^k}} {\left\{ \begin{array}{l}
{\bf{e}}_0^{}{\bf{e}}_0^T{\rho _{k + 1,n}} + \\
{\left[ \begin{array}{l}
{\bf{Q}}_k^{ - 1} - {\bf{Q}}_k^{ - 1}E\left\{ {{{\bf{F}}_k}} \right\} \times \\
{\left[ {{{\bf{J}}_{k,n}} + E\left\{ {{\bf{F}}_k^T{\bf{Q}}_k^{ - 1}{{\bf{F}}_k}} \right\}} \right]^{ - 1}}\\
 \times E\left\{ {{{\bf{F}}_k}} \right\}{\bf{Q}}_k^{ - 1}
\end{array} \right]^{ - 1}}\\
 \times \Pr \left( {{\Theta _{k,n}},{Z_{k + 1}} = \emptyset } \right)
\end{array} \right\}}  + \\
\sum\limits_{n = {2^k} + 1}^{{2^{k + 1}}} {\left\{ \begin{array}{l}
{\left[ \begin{array}{l}
{\bf{Q}}_k^{ - 1} + E\left\{ {{\bf{H}}_{k + 1}^T{\bf{R}}_{k + 1}^{ - 1}{\bf{H}}_{k + 1}^{}} \right\}\\
 - {\bf{Q}}_k^{ - 1}E\left\{ {{{\bf{F}}_k}} \right\} \times \\
{\left[ {{{\bf{J}}_{k,n - {2^k}}} + E\left\{ {{\bf{F}}_k^T{\bf{Q}}_k^{ - 1}{{\bf{F}}_k}} \right\}} \right]^{ - 1}}\\
 \times E\left\{ {{{\bf{F}}_k}} \right\}{\bf{Q}}_k^{ - 1}
\end{array} \right]^{ - 1}}\\
 \times \Pr \left( {{\Theta _{k,n - {2^k}}},{Z_{k + 1}} \ne \emptyset } \right)
\end{array} \right\}}
\end{array}
\end{equation}

It is obvious that (\ref{eq-d1}) is equal to (\ref{eq-d2}), in the condition that:
\begin{equation}\label{eq-d3}
\begin{array}{l}
 {\rho _{k + 1,n}} \\
  = \int { \cdots \int {f({X_{k + 1}} = \emptyset ,{\Theta _{k,n}},{Z_{k + 1,n}} = \emptyset )\delta {Z_{1,n}} \cdots \delta {Z_{k,n}}} }  \\
  = 0 \\
 \end{array}
\end{equation}

It means that ${P_{k + 1}}{\rm{(RFS)}}$ is equal to ${P_{k +
1}}(ENUM)$, when the target exists form the beginning to the end.

\section{Examples}\label{s6}
In this section, the previous theory is illustrated by two study cases: the one related to a linear filtering model and the other one referring to a non-linear bearings-only tracking.
\subsection{Linear Filtering Case}
This section illustrates the application of previous theoretical result by a linear case with Gaussian noise. When the target exists:
\begin{equation}\label{eq-44}
{X_k} = \left\{ {{{\bf{x}}_k}} \right\} = \left\{ {{{\left[ {\begin{array}{*{20}{c}}
   {{x_k}} & {{{\dot x}_k}} & {{y_k}} & {{{\dot y}_k}}  \\
\end{array}} \right]}^T}} \right\}
\end{equation}
The target is detected:
\begin{equation}\label{eq-45}
{Z_k} = \left\{ {{{\bf{z}}_k}} \right\} = \left\{ {{{\bf{z}}_k} =
{{\left[ {\begin{array}{*{20}{c}} {{z_{x,k}}}&{{z_{y,k}}}
\end{array}} \right]}^T}} \right\}
\end{equation}
The target motion is modeled as a linear equation:
\begin{equation}\label{eq-46}
\begin{array}{*{20}{c}}
{{{\bf{x}}_{k + 1}} = {{\bf{F}}_k}{{\bf{x}}_k} + {{\bf{w}}_k},}\\
{{{\bf{F}}_k} = \left[ {\begin{array}{*{20}{c}}
1&T&0&0\\
0&1&0&0\\
0&0&1&T\\
0&0&0&1
\end{array}} \right],}\\
{{{\bf{w}}_k} \sim N\left( {{\bf{0}},{{\bf{Q}}_k}} \right),}\\
{{{\bf{Q}}_k} = q\left[ {\begin{array}{*{20}{c}}
{{T^3}/3}&{{T^2}/2}&0&0\\
{{T^2}/2}&T&0&0\\
0&0&{{T^3}/3}&{{T^2}/2}\\
0&0&{{T^2}/2}&T
\end{array}} \right]}
\end{array}
\end{equation}

In this case, the time interval $T$ is 5s. The intensity of the process noise $q = {10^{ - 8}}$.

The measurement equation is as follow:
\begin{equation}\label{eq-47}
\begin{array}{*{20}{c}}
   {{{\bf{z}}_k} = {{\bf{H}}_k}{{\bf{x}}_k} + {{\bf{v}}_k},} & {{{\bf{H}}_k} = \left[ {\begin{array}{*{20}{c}}
   1 & 0 & 0 & 0  \\
   0 & 0 & 1 & 0  \\
\end{array}} \right],}  \\
   {{{\bf{v}}_k} \sim N\left( {{\bf{0}},{{\bf{R}}_k}} \right),} & {{{\bf{R}}_k} = diag\left( {\sigma _x^2,\sigma _y^2} \right)}  \\
\end{array}
\end{equation}

The variance of measurement noise is ${\sigma _x^2}={\sigma _y^2}=25^2$.

The initial target state standard variance is ${c_r} = 100,{c_v} = 5$, and the initial FIM is ${\bf{J}}_{0,1}^{ - 1} = diag(c_r^2,c_v^2,c_r^2,c_v^2)$.

The errors in cardinality mismatches are:
\begin{equation}\label{eq-48}
{\bf{e}}_1^{} = \left[ {\begin{array}{*{20}{c}}
   {{c_r}}  \\
   {{c_v}}  \\
\end{array}} \right],{\bf{e}}_0^{} = \left[ {\begin{array}{*{20}{c}}
   {{c_r}}  \\
   {{c_v}}  \\
\end{array}} \right]
\end{equation}
Fig.1, 2 and 3 show the root-mean-square error (RMSE) bounds between two sets   and   in (a) x-position and (b) x-velocity for ten scans. Similar results can be also obtained for the y-axis of position and velocity.

\subsubsection{The Influence of $r$}
In Fig.1, the unchanged parameters used in the computation of the bounds are ${P_d} = 0.8$ and $b = 1$. Here,   is the probability of the target existing initially. We compare the different bounds at various value of the parameter . $r$ is the maintenance probability. In addition, there is the   for case  using ENUM method as in (for short RMSE (ENUM)) \cite{Ref9}. When the target exist from the first time-step to the last one ($b = 1,r = 1$ for the models of RMSE (RFS)), this situation is the same as in which the RMSE (ENUM) is calculated. Moreover, RMSE (ENUM) is the true bound for the case of ${P_d} < 1$ and ${P_{fa}} = 0$.

In Fig.1, when $r = 1$, the RMSE (RFS) is approximate to the RMSE (ENUM), and as the number of the scans of measurements increases, they become the same. The reason is that, as discussed in section \ref{s5}, when it is not satisfied the condition that ${\bf{e}}_1^{}{\bf{e}}_1^T = {\bf{e}}_0^{}{\bf{e}}_0^T \gg {\left[ {{{\bf{J}}_{k,n}}} \right]^{ - 1}}$, RMSE (RFS) and RMSE (ENUM) could not be the same but just close. With the number of the scans of measurements increasing, it become satisfied that ${\bf{e}}_1^{}{\bf{e}}_1^T = {\bf{e}}_0^{}{\bf{e}}_0^T \gg {\left[ {{{\bf{J}}_{k,n}}} \right]^{ - 1}}$, and then RMSE (RFS) and RMSE (ENUM) are similar. This problem would be further discussed in the following passage.

When $0 \ll r < 1$, by taking account to the uncertain of the existence of the target, the bounds enhances. In reality, it means that, if the target disappeared with the probability $0.1$ ($r = 0.9,b = 1$), the performance of estimation would not reach the RMSE (ENUM), but the RMSE (RFS). In the other word, the calculation of RMSE (ENUM) is overly optimistic, when there is uncertainty about the target existing or not.

In addition, we confine ourselves to the matter under the situation $0 \ll r \le 1$ in the following discussion, when the initial probability of existing target $b = 1$. Because the case of $0 < r \ll 1, b = 1$ means that the target being preset or not is changed dramatically with time step. This is unrealistic and with little significance to discuss for tracking system.
\begin{figure}
  \centering
  \subfigure[Position in X-axis]{
    \label{fig:subfig:a} 
    \includegraphics[width=3in]{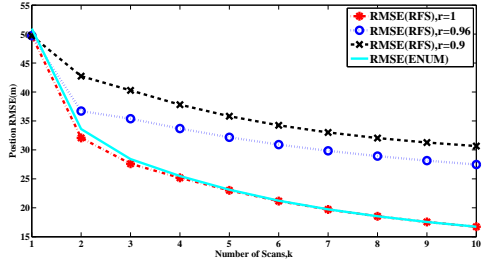}}
  \hspace{1in}
  \subfigure[Velocity in X-axis]{
    \label{fig:subfig:b} 
    \includegraphics[width=3in]{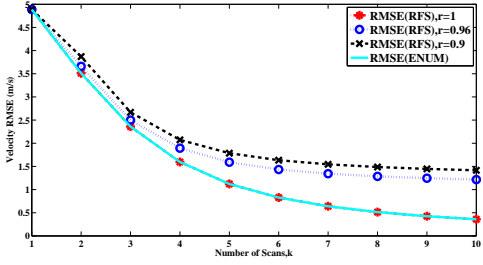}}
  \caption{Comparisons of RMSEs with different r (linear filtering)}
  \label{fig1} 
\end{figure}

\subsubsection{The Influence of $P_d$}
As shown in Fig.2, The RMSE (RFS) and RMSE (ENUM) are compared in the case of ${P_d} = 0.7$ and ${P_d} = 0.9$. The unchanged parameters are $r = 0.9$ and $b = 1$ for the models of RMSE (RFS). This means that the target enters at the first step, and then disappeared with the probability of $0.1$.

Fig.2 shows that the RMSE (RFS) is always larger than RMSE (ENUM), which is also illustrated in Fig.1. The reason is, when $r < 1$, the uncertainty of the existence of the target improves the bound of estimation. In Fig.2, when the probability of detection is reduced from $0.9$ to $0.7$, the RMSEs calculated by both the two methods are increased. However, the influence of the target appearance or disappearance is more significant than the influence of the miss detection. Because the error in cardinality mismatches has great effect in the calculation of the RMSE (RFS), which is paid no attention in the calculation of the RMSE (ENUM). Therefore, the RMSE (RFS) would be the true bound in the case of ${P_d} < 1$ and ${P_{fa}} = 0$, if the targets disappeared with certain probability.

\begin{figure}
  \centering
  \subfigure[Position in X-axis]{
    \label{fig:subfig:a} 
    \includegraphics[width=3in]{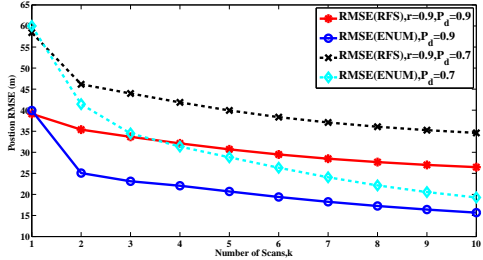}}
  \hspace{1in}
  \subfigure[Velocity in X-axis]{
    \label{fig:subfig:b} 
    \includegraphics[width=3in]{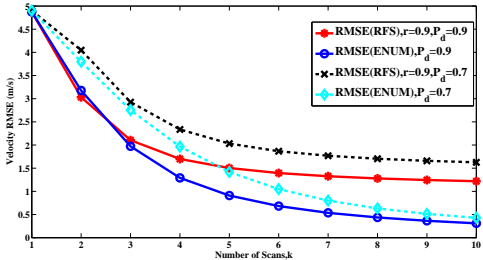}}
  \caption{Comparisons of RMSEs with different probability of detection (linear filtering)}
  \label{fig2} 
\end{figure}

\subsubsection{The Influence of Cardinality Mismatches}
In order to illustrate the discussion in \ref{s5}, we reset the errors of mismatches of cardinality ${{\bf{e}}_0}$ and ${{\bf{e}}_1}$:
\begin{equation}\label{eq-49}
{\bf{e}}_1^{} = \left[ {\begin{array}{*{20}{c}}
   {2{c_r}}  \\
   {2{c_v}}  \\
\end{array}} \right],{\bf{e}}_0^{} = \left[ {\begin{array}{*{20}{c}}
   {2{c_r}}  \\
   {2{c_v}}  \\
\end{array}} \right],
\end{equation}

\begin{figure}
  \centering
  \subfigure[Position in X-axis]{
    \label{fig:subfig:a} 
    \includegraphics[width=3in]{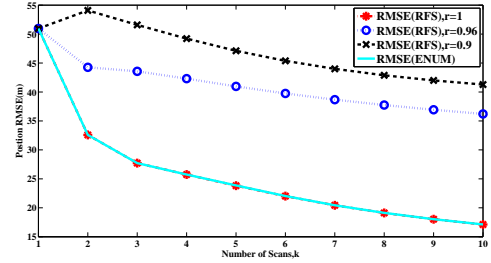}}
  \hspace{1in}
  \subfigure[Velocity in X-axis]{
    \label{fig:subfig:b} 
    \includegraphics[width=3in]{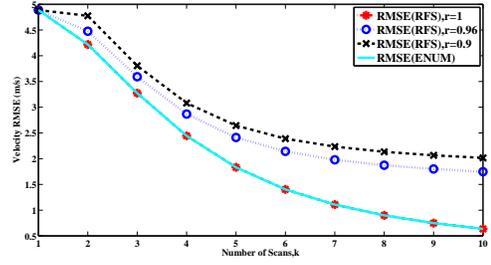}}
  \caption{Comparisons of RMSEs with different r with high error in cardinality mismatches (linear filtering)}
  \label{fig3} 
\end{figure}

The other settings are as similar as what in the Fig.1. Comparing
the Fig.1 and the Fig.3, when it is satisfied that
${\bf{e}}_1^{}{\bf{e}}_1^T = {\bf{e}}_0^{}{\bf{e}}_0^T \gg {\left[
{{{\bf{J}}_{k,n}}} \right]^{ - 1}}$, it is evident that the RMSE
(RFS) and RMSE (ENUM) are always the same.  As the enumeration PCRLB
is verified as the exact bound in the case of ${P_d} < 1$, our
method calculating tracking bound can always attain the optimal
method, when errors ${\bf{e}}_1^{}$ and ${\bf{e}}_0^{}$ are set as
(\ref{eq-49}). However, as in Fig.3, overestimated error in
cardinality mismatches leads to an unreasonable higher RMSE (RFS).
Moreover, the overrating RMSE (RFS) reduce slowly with the number of
the scans decreasing, which would be meaningless for tracking system
to some extent.  Therefore, the setting of the errors in cardinality
mismatches in (\ref{eq-48}) is more reasonable. As in figure.1, the
RMSE (RFS) and RMSE (ENUM) approach each other as the number of the
scans of measurements increases, and then they become the same at
last. In reality, it is usual that the errors bought by the
mismatches in the number of targets between the true state and
estimation, i.e. ${\bf{e}}_0^{}$ and ${\bf{e}}_1^{}$, are designed
according to initial FIM.

\subsection{Nonlinear Filtering Case}
This example is as similar as the bearings-only tracking case in \cite{Ref12}. This system can be applied in electro-magnetic (EM) equipment, electronic warfare devices (ESM) and passive sonar \cite{Ref12}.subsection text here.

The observer, named ownship, is a moving platform carrying sensor. Its state vector is denoted as ${\bf{x}}_k^o$ and assumed known. The target vector is denoted as ${\bf{x}}_k^t$. The relative state vector is defined as:
\begin{equation}\label{eq-50}
{{\bf{x}}_k} = {\bf{x}}_k^t - {\bf{x}}_k^o = {\left[ {\begin{array}{*{20}{c}}
   {{\chi _k}} & {{{\dot \chi }_k}} & {{\gamma _k}} & {{{\dot \gamma }_k}}  \\
\end{array}} \right]^T}
\end{equation}
where $\left( {{\chi _k},{\gamma _k}} \right)$ is the relative target position and $\left( {{{\dot \chi }_k},{{\dot \gamma }_k}} \right)$ is its velocity. The dynamic equation is as following:
\begin{equation}\label{eq-51}
{{\bf{x}}_{k + 1}} = {F_k}{{\bf{x}}_k} - {U_{k,k + 1}}
\end{equation}
where ${F_k}$ is denoted in (\ref{eq-46}), and the effect of a mismatch between the observer and the target motion model is accounted by :

\begin{equation}\label{eq-52}
{U_{k,k + 1}} = \left[ {\begin{array}{*{20}{c}}
   {\chi _{k + 1}^o - \chi _k^o - T\dot \chi _k^o}  \\
   {\dot \chi _{k + 1}^o - \dot \chi _k^o}  \\
   {\gamma _{k + 1}^o - \gamma _k^o - T\dot \gamma _k^o}  \\
   {\dot \gamma _{k + 1}^o - \dot \gamma _k^o}  \\
\end{array}} \right]
\end{equation}

The measurement equation is:
\begin{equation}\label{eq-53}
{z_k} = {h_k}\left( {{{\bf{x}}_k}} \right) + {v_k}
\end{equation}
where
\begin{equation}\label{eq-54}
{h_k}\left( {{{\bf{x}}_k}} \right) = \arctan \frac{{{\chi _k}}}{{{\gamma _k}}}
\end{equation}

${v_k}$ is a zero-mean white with covariance ${R_k} = \sigma _z^2 = {\left( {{1^ \circ }} \right)^2}$.  Based on (\ref{eq-6a}), the Jacobian of ${h_k}\left( {{{\bf{x}}_k}} \right)$ is calculated as:
\begin{equation}\label{eq-55}
{H_k} = \left[ {\begin{array}{*{20}{c}}
   {\frac{{{\gamma _k}}}{{\chi _k^2 + \gamma _k^2}}} & 0 & { - \frac{{{\chi _k}}}{{\chi _k^2 + \gamma _k^2}}} & 0  \\
\end{array}} \right].
\end{equation}

The initial target state standard variance is ${c_r} = 10000m,{c_v} = 100m/s$, and the initial FIM is ${\bf{J}}_{0,1}^{ - 1} = diag(c_r^2,c_v^2,c_r^2,c_v^2)$.

Ownship is moving as a uniform circular motion. The angular velocity is $\omega  = {1.0125^ \circ }/s$. The dynamic equation of the observer is given by:
\begin{equation}\label{eq-56}
{\bf{x}}_{k + 1}^o = {\Theta _k}{\bf{x}}_k^o
\end{equation}
\begin{equation}\label{eq-57}
{\Theta _k} = \left[ {\begin{array}{*{20}{c}}
   1 & {{\rm{sin(}}\omega {\rm{T)/}}\omega } & 0 & {{\rm{( - 1 + cos((}}\omega {\rm{T))/}}\omega }  \\
   0 & {{\rm{cos(}}\omega {\rm{T)}}} & 0 & {{\rm{ - sin(}}\omega {\rm{T)}}}  \\
   0 & {{\rm{(1 - cos(}}\omega {\rm{T))/}}\omega } & 1 & {{\rm{sin(}}\omega {\rm{T)/}}\omega }  \\
   0 & {{\rm{sin(}}\omega {\rm{T)}}} & 0 & {{\rm{cos(}}\omega {\rm{T)}}}  \\
\end{array}} \right]
\end{equation}

The initial target state vector $x_1^t = {\left[ {\begin{array}{*{20}{c}}{ - 25km} & {150m/s} & {20km} & {100m/s}  \\\end{array}} \right]^T}$ and the initial observer state vector $x_1^o = {\left[ {\begin{array}{*{20}{c}}{ - 30km} & {200m/s} & {50km} & {0m/s}  \\\end{array}} \right]^T}$.  The target and observer trajectories are shown in Fig.4.
\begin{figure}
    \includegraphics[width=3in]{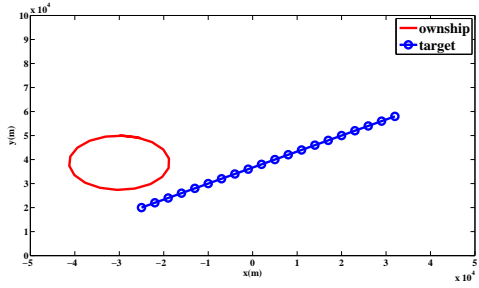}
  \caption{Bearing-only tracking scenario}
  \label{fig4} 
\end{figure}

Fig.5 and Fig.6 show the RMSE bound between two sets ${X_k}$ and ${\hat X_k}({Z_1} \cdots {Z_k})$ in (a) y-position and (b) y-velocity for twenty scans. Similar results can be also obtained for the x-axis of position and velocity.

\subsubsection{The Influence of $r$}
In Fig.5, the unchanged parameters used in the computation of the bounds are   and  . We concentrate on the influence of various value of the parameter . Since there is no process noise for both ownship and target, the calculation of FIM is based on (\ref{eq-36}).

As shown in Fig.5, $b = 1,r = 1$ for the models of RMSE (RFS) mean that the probability of  target existence is unity. This condition can be compare with that of calculating the RMSE (ENUM) for they are the same scenario.

Note that, in this bearings-only tracking case (Fig.5), the RMSE decrease more steeply than which in the linear case (Fig.1). The reason is, at the initial several scans, initial FIM ${\bf{J}}_{0,1}^{ - 1}$ impacts the RMSE most, because the measurement of target is missed initially with high probability, comparing to the following scans. However, the covariance matrix of measurement noise ${{\bf{R}}_k}$ influences the estimated error more with more measurements observed.  In the Fig.5, there is the relationship that ${\bf{J}}_{0,1}^{ - 1} \gg {{\bf{R}}_k}$, while in the Fig.1, the ${\bf{J}}_{0,1}^{ - 1}$ is bigger than ${{\bf{R}}_k}$, but closely. Therefore, the bounds shown in Fig.5 is influenced by the initial FIM ${\bf{J}}_{0,1}^{ - 1}$ more than which in Fig.1.

Moreover, in the Fig.5, when $r < 1$, the bounds intersect the bound
of $r = 1$. While, in the Fig.1, these bounds always exceed the
bound of $r = 1$.  The reasons are not only the initial FIM
${\bf{J}}_{0,1}^{ - 1} \gg {{\bf{R}}_k}$ discussed above, but also
the error ${\bf{e}}\left( {X = \emptyset, \hat X\left( Z \right) =
\emptyset } \right) = {\bf{0}}$, which is defined in (\ref{eq-14}).
The initial target state standard variance is ${c_r} = 100m$ in the
linear filtering case, while which in the nonlinear one is ${c_r} =
10000m$. Therefore, the estimation of absence of target, where
${\bf{e}}\left( {X = \emptyset, \hat X\left( Z \right) = \emptyset }
\right) = {\bf{0}}$, plays more roles in the bounds when $r < 1$ in
the Fig.5, and contributes that the bound of $r < 1$ is lower than
that of  $r = 1$. In fact, by the accumulation of several scans, the
bounds $r < 1$ must be no lower than that of $r = 1$, which is also
shown in the Fig.5, because it take the uncertainty of the existence
of target.

\begin{figure}
  \centering
  \subfigure[Position in Y-axis]{
    \label{fig:subfig:a} 
    \includegraphics[width=3in]{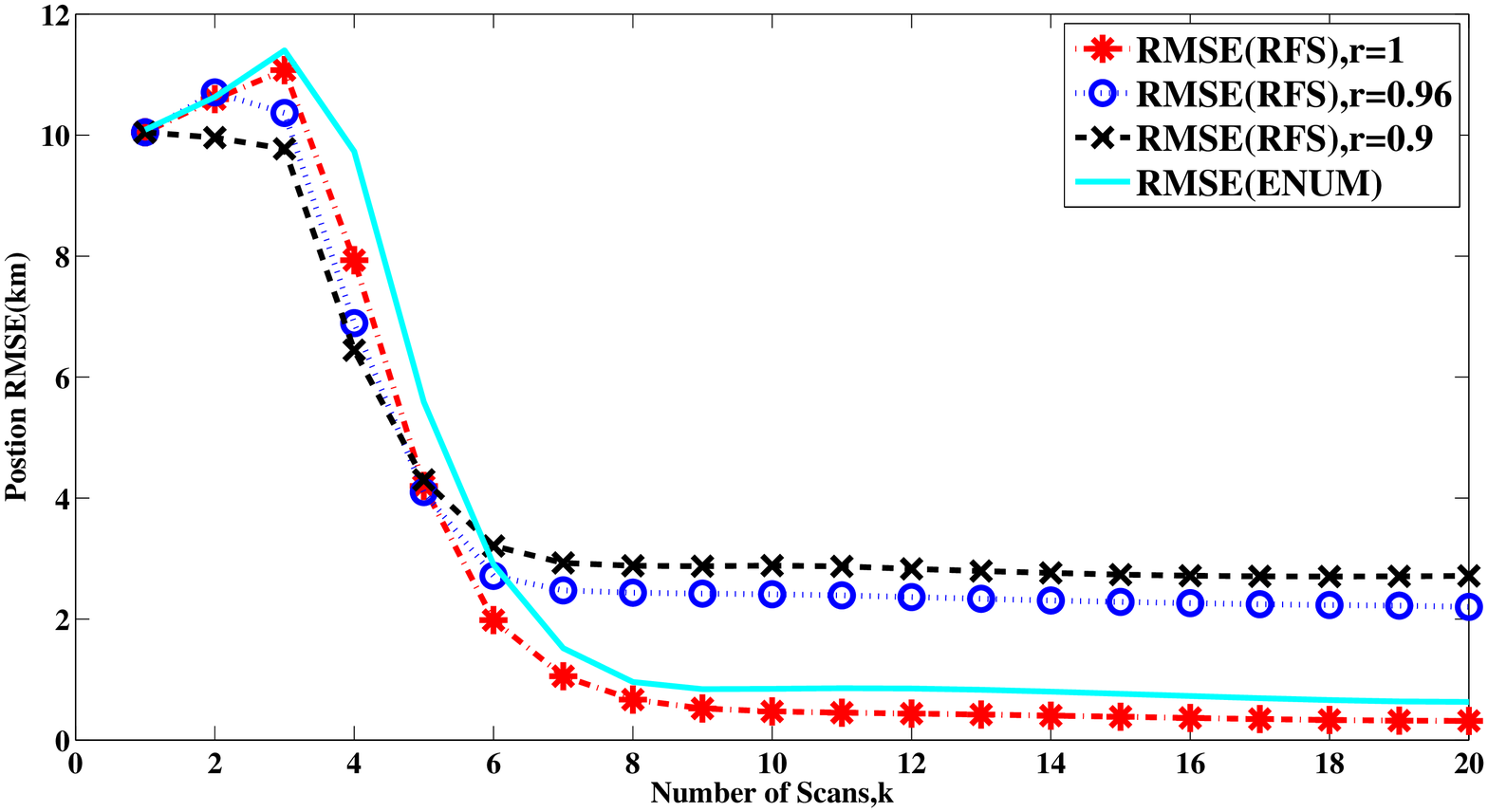}}
  \hspace{1in}
  \subfigure[Velocity in Y-axis]{
    \label{fig:subfig:b} 
    \includegraphics[width=3in]{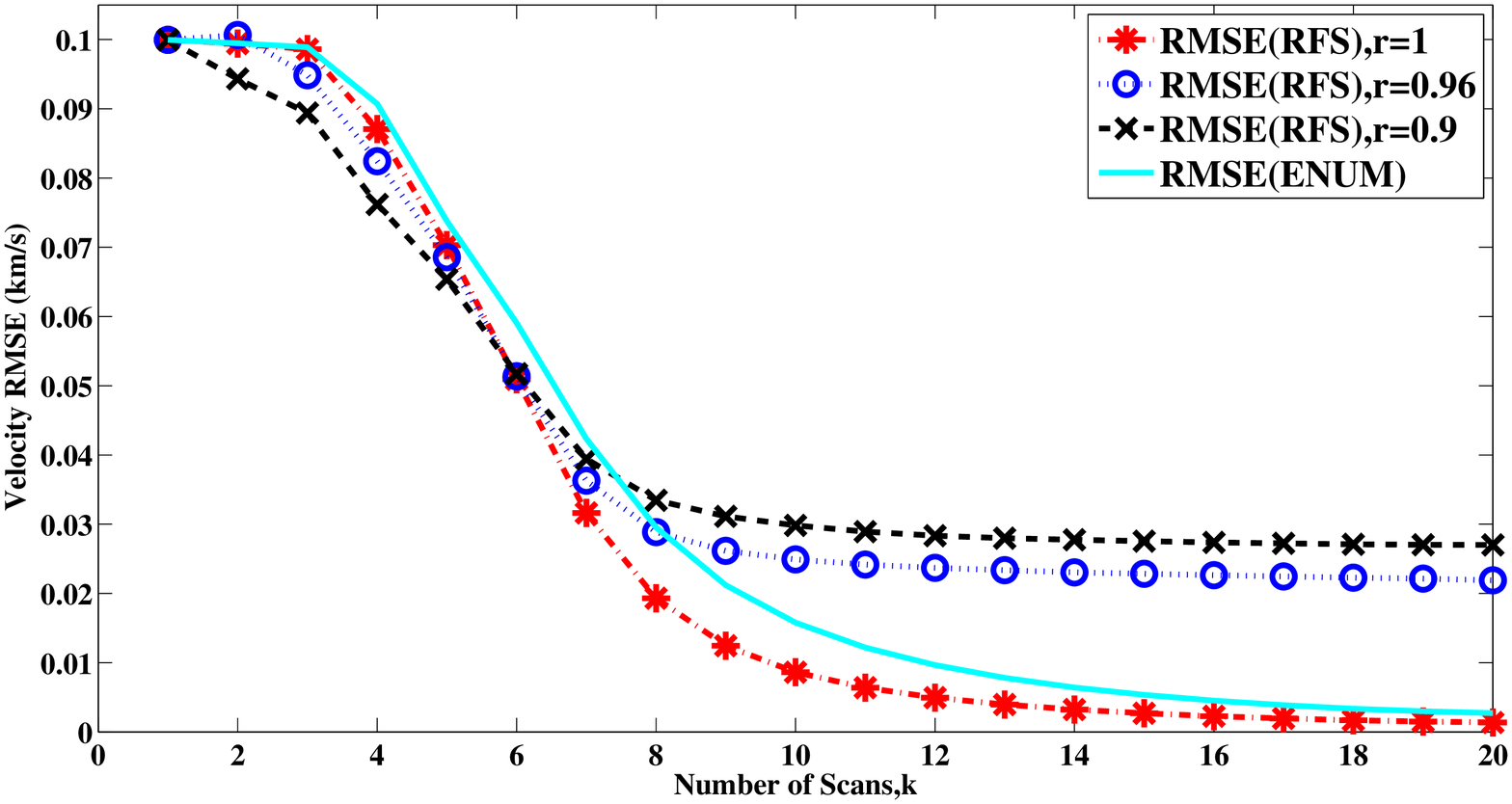}}
  \caption{Comparisons of RMSEs with different r (nonlinear filtering)}
  \label{fig5} 
\end{figure}
\subsubsection{The Influence of $b$}

In order to indicate the influence of the target existing or not
initially, the value of $b$ is changed. $b$ is the probability of
the target existing initially. In all previous examples, we consider
the situations that target exist at the first step ($b = 1$). In the
Fig.6, we reset $b = 0.1$, which means that the target appear on the
probability of $0.1$ at the first time step. In this case, the
bounds of RMSE (RFS) is hard to compare with the RMSE (ENUM),
because when the RMSE (ENUM) is calculated, it is the same situation
that $b = 1,r = 1$for the calculation of RMSE (RFS). Nevertheless,
the relationship between RMSE (ENUM) and RMSE (RFS) is discussed in
detail in the linear example and the need of comparing them is
little in this example. Here we contain $b = 0.1$ and vary $r$ in
the Fig.6. In Fig.6, the probability of detection is set ${P_d} =
0.9$ for all bounds.

As shown in Fig.6, the influence of ${\bf{e}}\left( {X = \emptyset,
\hat X\left( Z \right) = \emptyset } \right) = {\bf{0}}$ is more
significant than Fig.5. Because this case means that at first step,
the state set of the target is empty with the probability $0.9$, and
then the cardinality turn to one with the probability $0.1$ ($r =
0.9,b = 0.1$). In the other word, the target enter with the probably
$1-r$.  When $r < 1$, the bound of RMSE (RFS) is bigger than RMSE
(ENUM) in Fig.6, except the initial several scans. The optimality of
the RMSE (RFS) is verified again. It is noted that, in the case of ,
the bound of RMSE (RFS) with $r=1$ is always less than RMSE (ENUM).
This means that the target births at the first time step with the
probability 0.1, and keep the similar state as the first scan. As in
Fig.6, to the situation that the target is rare, the estimation of
empty set can reduce the error.

\begin{figure}
  \centering
  \subfigure[Position in Y-axis]{
    \label{fig:subfig:a} 
    \includegraphics[width=3in]{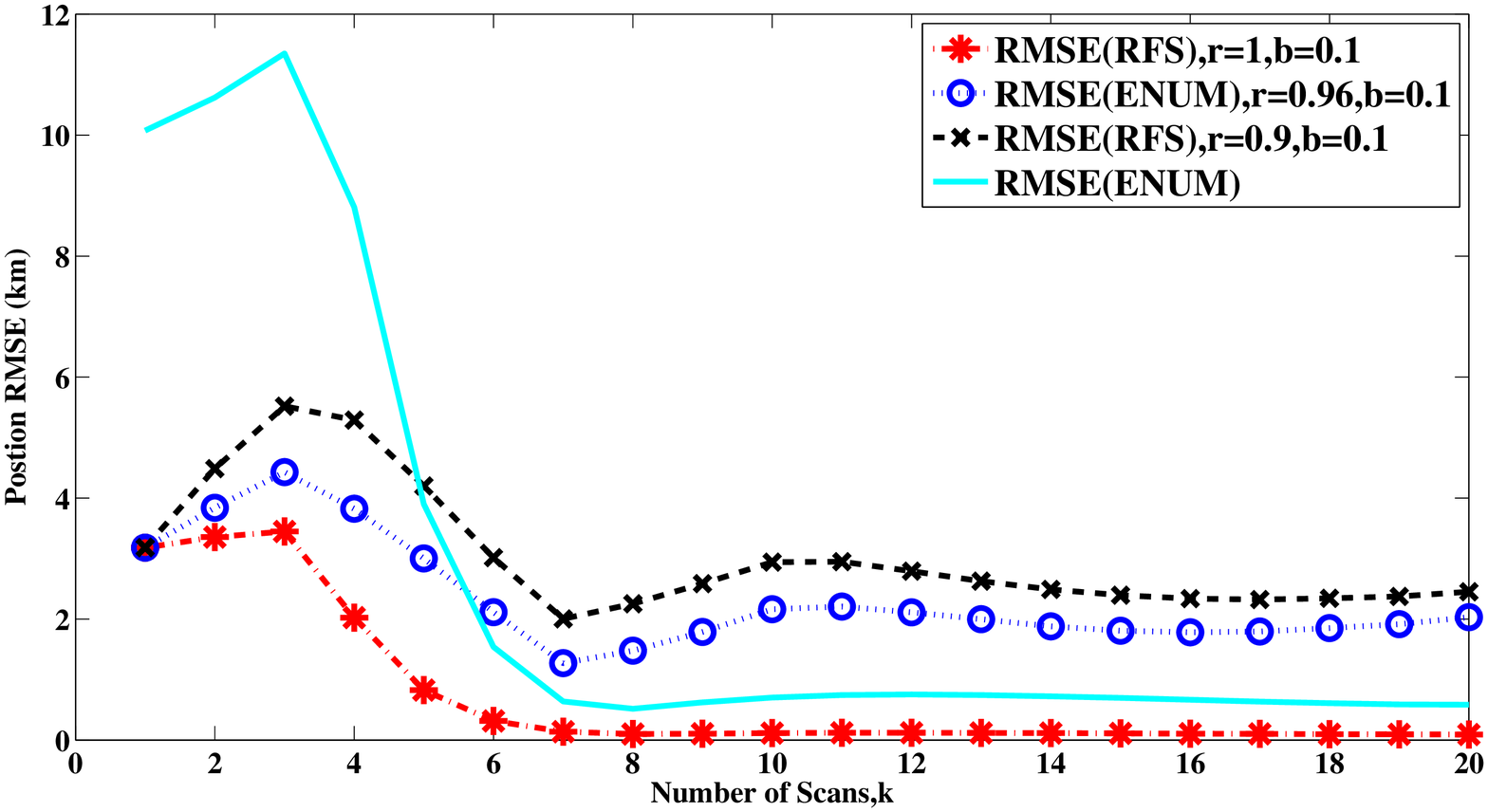}}
  \hspace{1in}
  \subfigure[Velocity in Y-axis]{
    \label{fig:subfig:b} 
    \includegraphics[width=3in]{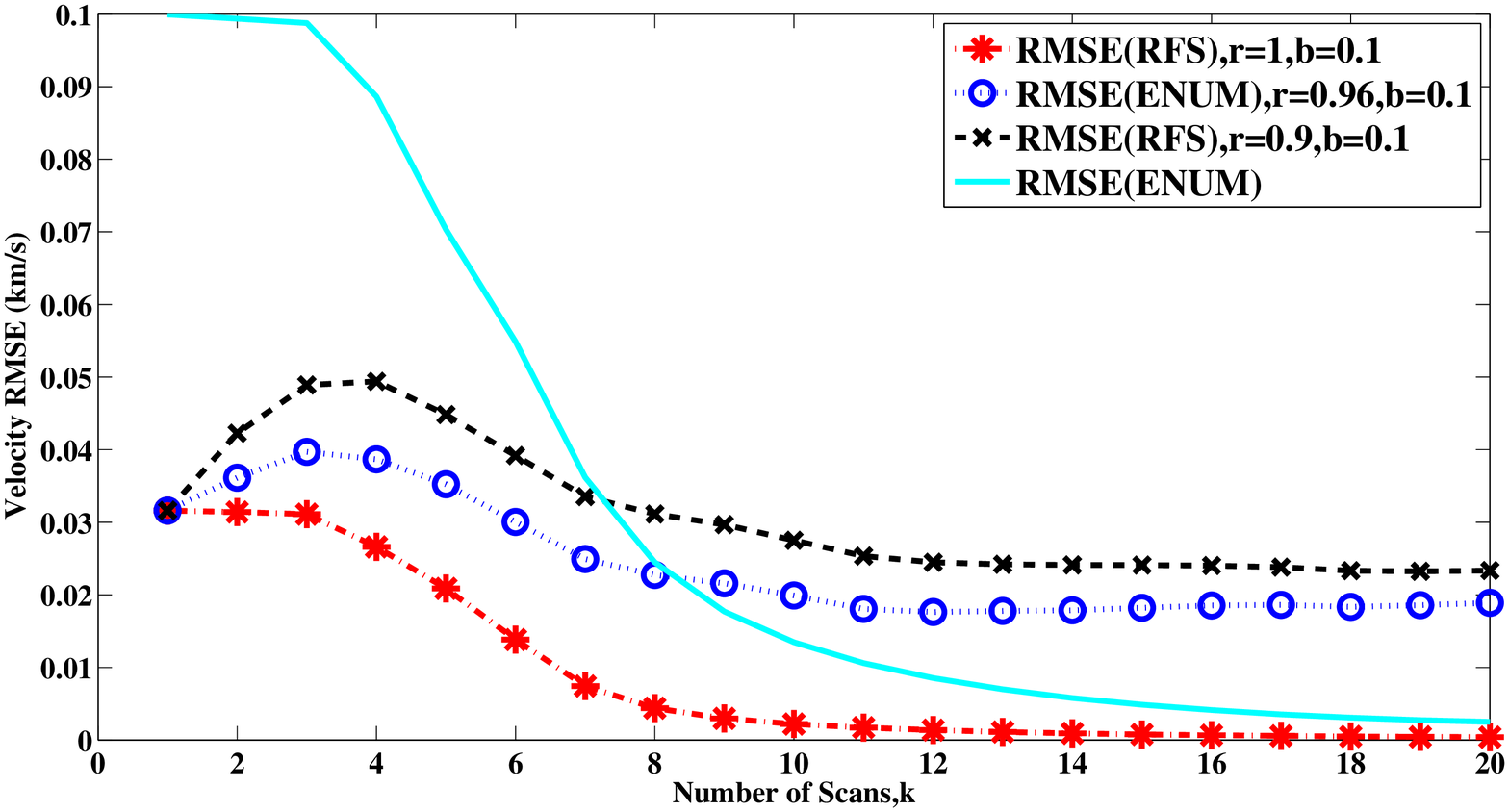}}
  \caption{Comparisons of RMSEs with b=0.1 and varying r (nonlinear filtering)}
  \label{fig65} 
\end{figure}

\section{Mathematic Proofs}\label{s7}
\subsection{Proof of Proposition 1}
When ${Z_{k + 1}} \ne \emptyset $, the error bound at time-step $k + 1$ is as following:
\begin{equation}\label{eq-58}
\begin{array}{l}
 {\Sigma _{k + 1,n}} = \int { \cdots \int \begin{array}{l}
 {{\bf{C}}_{k + 1,n}}*p({X_{k + 1}},{\Theta _{k + 1,n - {2^k}}},{Z_{k + 1,n}}) \\
 \delta {X_{k + 1}}\delta {Z_{1,n}} \cdots \delta {Z_{k + 1,n}} \\
 \end{array} }  \\
  = \int { \cdots \int \begin{array}{l}
 \left( {{{\bf{x}}_{k + 1}} - {{{\bf{\hat x}}}_{k + 1}}} \right){\left( {{{\bf{x}}_{k + 1}} - {{{\bf{\hat x}}}_{k + 1}}} \right)^T}* \\
 *p({X_{k + 1,n}} = \left\{ {{{\bf{x}}_{k + 1}}} \right\},{\Theta _{k,n - {2^k}}},{Z_{k + 1,n}} = \left\{ {{{\bf{z}}_{k + 1}}} \right\}) \\
 d{{\bf{x}}_{k + 1}}\delta {Z_{1,n}} \cdots \delta {Z_{k,n}}d{{\bf{z}}_{k + 1}} \\
 \end{array} }  \\
  \ge \left[ \begin{array}{l}
 {\bf{Q}}_k^{ - 1} + E\left\{ {{\bf{H}}_{k + 1}^T{\bf{R}}_{k + 1}^{ - 1}{\bf{H}}_{k + 1}^{}} \right\} -  \\
 {\bf{Q}}_k^{ - 1}E\left\{ {{{\bf{F}}_k}} \right\}{\left[ {{{\bf{J}}_k} + E\left\{ {{\bf{F}}_k^T{\bf{Q}}_k^{ - 1}{{\bf{F}}_k}} \right\}} \right]^{ - 1}}E\left\{ {{{\bf{F}}_k}} \right\}{\bf{Q}}_k^{ - 1} \\
 \end{array} \right] \\
 \begin{array}{*{20}{c}}
   {} & {}  \\
\end{array}*\Pr \left( {{\Theta _{k + 1,n}}} \right) \\
  = {\left[ {{{\bf{J}}_{k + 1,n}}} \right]^{ - 1}}*\Pr \left( {{\Theta _{k,n - {2^k}}},{Z_{k + 1}} \ne \emptyset } \right){\rm{ }} \\
 \end{array}
\end{equation}

When ${Z_{k + 1}} = \emptyset $, the error bound is given follow:
\begin{equation}\label{eq-59}
\begin{array}{l}
 {\Sigma _{k + 1,n}} \\
  = \int { \cdots \int \begin{array}{l}
 {{\bf{C}}_{k + 1,n}}*f({X_{k + 1}},{\Theta _{k,n}},{Z_{k + 1,n}} = \emptyset ) \\
 \delta {X_{k + 1}}\delta {Z_{1,n}} \cdots \delta {Z_{k,n}} \\
 \end{array} }  \\
  = \int { \cdots \int \begin{array}{l}
 {\bf{e}}\left( {\emptyset ,{{\hat X}_{k + 1}}({\Theta _{k,n}},{Z_{k + 1,n}} = \emptyset )} \right)* \\
 {\bf{e}}{\left( {\emptyset ,{{\hat X}_{k + 1}}({\Theta _{k,n}},{Z_{k + 1,n}} = \emptyset )} \right)^T}* \\
 p({X_{k + 1}} = \emptyset ,{\Theta _{k,n}},{Z_{k + 1,n}} = \emptyset )\delta {Z_{1,n}} \cdots \delta {Z_{k,n}} \\
 \end{array} }  \\
  + \int { \cdots \int \begin{array}{l}
 {\bf{e}}\left( {{X_{k + 1}} = \left\{ {{{\bf{x}}_{k + 1}}} \right\},{{\hat X}_{k + 1}}({\Theta _{k,n}},{Z_{k + 1,n}} = \emptyset )} \right)* \\
 {\bf{e}}{\left( {{X_{k + 1}} = \left\{ {{{\bf{x}}_{k + 1}}} \right\},{{\hat X}_{k + 1}}({\Theta _{k,n}},{Z_{k + 1,n}} = \emptyset )} \right)^T}* \\
 f\left( {{X_{k + 1}} = \left\{ {{{\bf{x}}_{k + 1}}} \right\},{\Theta _{k,n}},{Z_{k + 1,n}} = \emptyset } \right) \\
 d{{\bf{x}}_{k + 1}}\delta {Z_{1,n}} \cdots \delta {Z_{k,n}} \\
 \end{array} }  \\
 \end{array}
\end{equation}
If  ${\hat X_{k + 1}}({\Theta _{k,n}},{Z_{k + 1,n}} = \emptyset ) =
\emptyset $, take (\ref{eq-33}) into the (\ref{eq-59}):
\begin{equation}\label{eq-60}
\begin{array}{l}
 {\Sigma _{k + 1,n}} = {\bf{e}}_1^{}{\bf{e}}_1^T\int { \cdots \int \begin{array}{l}
 p({X_{k + 1}} = \left\{ {{x_{k + 1}}} \right\},{\Theta _{k,n}},{Z_{k + 1,n}} = \emptyset ) \\
 d{{\bf{x}}_{k + 1}}\delta {Z_{1,n}} \cdots \delta {Z_{k,n}} \\
 \end{array} }  \\
  = {\bf{e}}_1^{}{\bf{e}}_1^T\left( {\Pr \left( {{\Theta _{k,n}},{Z_{k + 1,n}} = \emptyset } \right) - {\rho _{k + 1,n}}} \right) \\
  = {\bf{P}}_{k + 1,n}^* \\
 \end{array}
\end{equation}

If ${\hat X_{k + 1}}({\Theta _{k,n}},{Z_{k + 1,n}} = \emptyset ) = {{\bf{\tilde x}}_{k + 1}}$, then the (\ref{eq-59}) reduces to:
\begin{equation}\label{eq-61}
\begin{array}{l}
 {\Sigma _{k + 1,n}} = {\bf{e}}_0^{}{\bf{e}}_0^T\int { \cdots \int \begin{array}{l}
 p({X_{k + 1}} = \emptyset ,{\Theta _{k,n}},{Z_{k + 1,n}}{\rm{ = }}\emptyset ) \\
 \delta {Z_{1,n}} \cdots \delta {Z_{k,n}} \\
 \end{array} }  \\
  + \int { \cdots \int \begin{array}{l}
 \left( {{{\bf{x}}_{k + 1}} - {{{\bf{\tilde x}}}_{k + 1}}} \right){\left( {{{\bf{x}}_{k + 1}} - {{{\bf{\tilde x}}}_{k + 1}}} \right)^T}p({X_{k + 1}} = \left\{ {{{\bf{x}}_{k + 1}}} \right\} \\
 ,{\Theta _{k,n}},{Z_{k + 1,n}}{\rm{ = }}\emptyset )d{{\bf{x}}_{k + 1}}\delta {Z_{1,n}} \cdots \delta {Z_{k,n}} \\
 \end{array} }  \\
  \ge {\bf{e}}_0^{}{\bf{e}}_0^T{\rho _{k + 1,n}} +  \\
 {\left[ {{\bf{Q}}_k^{ - 1} - {\bf{Q}}_k^{ - 1}E\left\{ {{{\bf{F}}_k}} \right\}{{\left[ {{{\bf{J}}_k} + E\left\{ {{\bf{F}}_k^T{\bf{Q}}_k^{ - 1}{{\bf{F}}_k}} \right\}} \right]}^{ - 1}}E\left\{ {{{\bf{F}}_k}} \right\}{\bf{Q}}_k^{ - 1}} \right]^{ - 1}}* \\
 \Pr \left( {{\Theta _{k,n}},{Z_{k + 1}} = \emptyset } \right) \\
  = {\bf{e}}_0^{}{\bf{e}}_0^T{\rho _{k + 1,n}} + {\left[ {{{\bf{J}}_{k + 1,n}}} \right]^{ - 1}}*\Pr \left( {{\Theta _{k,n}},{Z_{k + 1}} = \emptyset } \right) \\
  = {\bf{P}}_{k + 1,n}^{**} \\
 \end{array}
\end{equation}

The lower bound will be the minimum of the bounds of (\ref{eq-60}) and (\ref{eq-61}) (the proof detailed to one step MSE in \cite{Ref13}).

\subsection{Proof of Proposition 2}
From the definition of conditional probability, the relationship between $\Pr \left( {{\Theta _{k + 1,n}}} \right)$ and $\Pr \left( {{\Theta _{k,n}}} \right)$ is as follow:
\begin{equation}\label{eq-62}
\Pr \left( {{\Theta _{k + 1,n}}} \right) = \Pr \left( {{\Theta _{k,n}}} \right)p\left( {{Z_{k + 1,n}}\left| {{\Theta _{k,n}}} \right.} \right)
\end{equation}

It is obvious that we should determine the recursion of $p\left( {{Z_{k + 1,n}}\left| {{\Theta _{k,n}}} \right.} \right)$.

When the sequence number $n$ is in the range that ${2^k} + {2^{k - 1}} + 1 \le n \le {2^{k + 1}}$, the measurement ${Z_{k,n}} \ne \emptyset$ ${Z_{k+1,n}} \ne \emptyset$. It means that the target exists at time step $k$, still exists at time step $k+1$, and is observed. From the dynamical and measurement model, we can see that

\begin{equation}\label{eq-63}
p\left( {{Z_{k + 1,n}} \ne \emptyset \left| {{\Theta _{k,n - {2^k}}}} \right.} \right) = r{P_d},{2^k} + {2^{k - 1}} + 1 \le n \le {2^{k + 1}}
\end{equation}

and
\begin{equation}\label{eq-64}
p\left( {{Z_{k + 1,n}} = \emptyset \left| {{\Theta _{k,n}}} \right.} \right) = 1 - r{P_d},{2^{k - 1}} + 1 \le n \le {2^k}
\end{equation}

When the measurement ${Z_{k,n}} \ne \emptyset$, the state of the target is uncertain. Furthermore, it is hard to determine the state at time step $k+1$ and whether the measurement is empty or not. Because there is the relationship that: $p\left( {{Z_{k + 1,n}} \ne \emptyset \left| {{\Theta _{k,n}}} \right.} \right) = 1 - p\left( {{Z_{k + 1,n}} = \emptyset \left| {{\Theta _{k,n}}} \right.} \right)$. Therefore, the key is to get the recursive form of $p\left( {{Z_{k + 1,n}} = \emptyset \left| {{\Theta _{k,n}}} \right.} \right)$. This is discussed in the proposition 4.

\subsection{Proof of Proposition 3}
As in \cite{Ref14}, the recursive Bayes filter for RFS tracking system:
\begin{equation}\label{eq-65}
p\left( {{X_k}\left| {{\Theta _{k - 1,n}}} \right.} \right) = \int {f\left( {{X_k}\left| {{X_{k - 1}}} \right.} \right)} p\left( {{X_{k - 1}}\left| {{\Theta _{k - 1,n}}} \right.} \right)\delta {X_{k - 1}}
\end{equation}
\begin{equation}\label{eq-66}
p\left( {{X_k}\left| {{\Theta _{k - 1,n}},{Z_{k,n}}} \right.} \right) = \frac{{g\left( {{Z_{k,n}}\left| {{X_k}} \right.} \right)p\left( {{X_k}\left| {{\Theta _{k - 1,n}}} \right.} \right)}}{{p\left( {{Z_{k,n}}\left| {{\Theta _{k - 1,n}}} \right.} \right)}}
\end{equation}
\begin{equation}\label{eq-67}
p\left( {{Z_{k,n}}\left| {{\Theta _{k - 1,n}}} \right.} \right) = \int {g\left( {{Z_{k,n}}\left| {{X_k}} \right.} \right)p\left( {{X_k}\left| {{\Theta _{k - 1,n}}} \right.} \right)\delta {X_k}}
\end{equation}
Extending (\ref{eq-67}) by the measurement model (\ref{eq-27}) and (\ref{eq-28}):
\begin{equation}\label{eq-68}
\begin{array}{l}
 p\left( {{Z_{k,n}} = \emptyset \left| {{\Theta _{k - 1,n}}} \right.} \right) \\
  = g\left( {{Z_{k,n}} = \emptyset \left| {{X_k} = \emptyset } \right.} \right)p\left( {{X_k} = \emptyset \left| {{\Theta _{k - 1,n}}} \right.} \right) \\
  + \int {g\left( {{Z_{k,n}} = \emptyset \left| {{X_k} = \left\{ {{{\bf{x}}_k}} \right\}} \right.} \right)p\left( {{X_k} = \left\{ {{{\bf{x}}_k}} \right\}\left| {{\Theta _{k - 1,n}}} \right.} \right)d{{\bf{x}}_k}}  \\
  = p\left( {{X_k} = \emptyset \left| {{\Theta _{k - 1,n}}} \right.} \right) + \left( {1 - {P_d}} \right)\left( {1 - p\left( {{X_k} = \emptyset \left| {{\Theta _{k - 1,n}}} \right.} \right)} \right) \\
  = \left( {1 - {P_d}} \right) + {P_d}p\left( {{X_k} = \emptyset \left| {{\Theta _{k - 1,n}}} \right.} \right) \\
 \end{array}
\end{equation}

Hence (\ref{eq-66}) reduces to:
\begin{equation}\label{eq-69}
\begin{array}{l}
 p\left( {{X_k} = \emptyset \left| {{\Theta _{k - 1,n}},{Z_{k,n}} = \emptyset } \right.} \right) \\
  = \frac{{g\left( {{Z_{k,n}} = \emptyset \left| {{X_k} = \emptyset } \right.} \right)p\left( {{X_k} = \emptyset \left| {{\Theta _{k - 1,n}}} \right.} \right)}}{{p\left( {{Z_{k,n}} = \emptyset \left| {{\Theta _{k - 1,n}}} \right.} \right)}} \\
  = \frac{{p\left( {{X_k} = \emptyset \left| {{\Theta _{k - 1,n}}} \right.} \right)}}{{\left( {1 - {P_d}} \right) + {P_d}p\left( {{X_k} = \emptyset \left| {{\Theta _{k - 1,n}}} \right.} \right)}} \\
  = \frac{{p\left( {{Z_{k,n}} = \emptyset \left| {{\Theta _{k - 1,n}}} \right.} \right) - \left( {1 - {P_d}} \right)}}{{{P_d}p\left( {{Z_{k,n}} = \emptyset \left| {{\Theta _{k - 1,n}}} \right.} \right)}} \\
 \end{array}
\end{equation}

Therefore, the recursion of ${\rho _{k,n}}$ is given following:
\begin{equation}\label{eq-70}
\begin{array}{l}
 {\rho _{k + 1,n}} \\
  = \int { \cdots \int {p({X_{k + 1}} = \emptyset ,{\Theta _{k,n}},{Z_{k + 1,n}} = \emptyset )\delta {Z_{1,n}} \cdots \delta {Z_{k,n}}} }  \\
  = \int { \cdots \int \begin{array}{l}
 p\left( {{X_{k + 1}} = \emptyset \left| {{\Theta _{k,n}},{Z_{k + 1,n}} = \emptyset } \right.} \right)* \\
 p\left( {{\Theta _{k,n}},{Z_{k + 1,n}} = \emptyset } \right)\delta {Z_{1,n}} \cdots \delta {Z_{k,n}} \\
 \end{array} }  \\
  = \int { \cdots \int \begin{array}{l}
 \frac{{p\left( {{Z_{k + 1,n}} = \emptyset \left| {{\Theta _{k,n}}} \right.} \right) - \left( {1 - {P_d}} \right)}}{{{P_d}p\left( {{Z_{k + 1,n}} = \emptyset \left| {{\Theta _{k,n}}} \right.} \right)}}p\left( {{Z_{k + 1,n}} = \emptyset \left| {{\Theta _{k,n}}} \right.} \right) \\
 *\Pr \left( {{\Theta _{k,n}}} \right)\delta {Z_{1,n}} \cdots \delta {Z_{k,n}} \\
 \end{array} }  \\
  = \Pr ({\Theta _{k,n}})\int { \cdots \int {\frac{{p\left( {{Z_{k + 1,n}} = \emptyset \left| {{\Theta _{k,n}}} \right.} \right) - \left( {1 - {P_d}} \right)}}{{{P_d}}}} } \delta {Z_{1,n}} \cdots \delta {Z_{k,n}} \\
 \end{array}
\end{equation}

\subsection{Proof of Proposition 4}
According to the dynamical model (\ref{eq-24}) and (\ref{eq-25}), (\ref{eq-65}) turns into:
\begin{equation}\label{eq-71}
\begin{array}{l}
 p\left( {{X_k} = \emptyset \left| {{\Theta _{k - 1,n}}} \right.} \right) \\
  = p\left( {{X_k} = \emptyset \left| {{X_{k - 1}} = \emptyset } \right.} \right)p\left( {{X_{k - 1}} = \emptyset \left| {{\Theta _{k - 1,n}}} \right.} \right) +  \\
 \int \begin{array}{l}
 p\left( {{X_k} = \emptyset \left| {{X_{k - 1}} = \left\{ {{{\bf{x}}_{k - 1}}} \right\}} \right.} \right)* \\
 p\left( {{X_{k - 1}} = \left\{ {{{\bf{x}}_{k - 1}}} \right\}\left| {{\Theta _{k - 1,n}}} \right.} \right)d{{\bf{x}}_{k - 1}} \\
 \end{array}  \\
  = rp\left( {{X_{k - 1}} = \emptyset \left| {{\Theta _{k - 1,n}}} \right.} \right) +  \\
 \left( {1 - r} \right)\left( {1 - p\left( {{X_{k - 1}} = \emptyset \left| {{\Theta _{k - 1,n}}} \right.} \right)} \right) \\
  = \left( {1 - r} \right) + \left( {2r - 1} \right)p\left( {{X_{k - 1}} = \emptyset \left| {{\Theta _{k - 1,n}}} \right.} \right) \\
 \end{array}
\end{equation}

According to (\ref{eq-68}), (\ref{eq-69}) and (\ref{eq-71}), if $1 \le n \le {2^{k - 1}}$, the conditional probability of ${Z_{k + 1,n}} = \emptyset$ is derived as follow:
\begin{equation}\label{eq-72}
\begin{array}{l}
 p\left( {{Z_{k + 1,n}} = \emptyset \left| {{\Theta _{k,n}}} \right.} \right) \\
  = p\left( {{Z_{k + 1,n}} = \emptyset \left| {{\Theta _{k - 1,n}},{Z_{k,n}} = \emptyset } \right.} \right) \\
  = \left( {1 - {P_d}} \right) + {P_d}p\left( {{X_{k + 1}} = \emptyset \left| {{\Theta _{k - 1,n}},{Z_{k,n}} = \emptyset } \right.} \right) \\
  = 1 - r{P_d} + {P_d}\left( {2r - 1} \right)p\left( {{X_k} = \emptyset \left| {{\Theta _{k - 1,n}},{Z_{k,n}} = \emptyset } \right.} \right) \\
  = 1 - r{P_d} + {P_d}\left( {2r - 1} \right)\frac{{p\left( {{Z_{k,n}} = \emptyset \left| {{\Theta _{k - 1,n}}} \right.} \right) - \left( {1 - {P_d}} \right)}}{{{P_d}p\left( {{Z_{k,n}} = \emptyset \left| {{\Theta _{k - 1,n}}} \right.} \right)}} \\
 \end{array}
\end{equation}

\section{Conclusion}\label{s8}
In this paper, a performance bound for dynamic estimation and
filtering problem, in the framework of finite set statistics, is
presented for the first time. This bound is recursion, and hence it
is significant for performance evaluation of tracking systems. In
addition, the case of $P_d<1$ is taken into account, which makes
this bound realistically. Moreover, this bound shows the influence
of the uncertainty of target existence.

The discussion and numerical examples show that our bound can obtain
the enumeration PCRLB, which is the true bound in the case of
$P_d<1$ and $P_{FA}=0$, when the target remains form the beginning
to the end. Furthermore, for some targets of high uncertainty, which
may appear or disappearance with certain probability, our bound is
more accurate and reasonable than enumeration PCRLB. To this
situation, by considering whether the state set of the target is
empty or not, the bound calculated in this paper is more general
than previous bounds in the framework of random vector statistics.


%

\appendices




\ifCLASSOPTIONcaptionsoff
  \newpage
\fi

\end{document}